# Analytical models for coated plasmonic particles: effects of shape and size-corrected dielectric function

Nikolai G. Khlebtsov[1,2,*], Sergey V. Zarkov[1,3]

[1]Institute of Biochemistry and Physiology of Plants and Microorganisms, "Saratov Scientific Centre of the Russian Academy of Sciences," 13 Prospekt Entuziastov, Saratov 410049, Russia

[2]Saratov State University, 83 Ulitsa Astrakhanskaya, Saratov 410012, Russia

[3]Institute of Precision Mechanics and Control, "Saratov Scientific Centre of the Russian Academy of Sciences," 24 Ulitsa Rabochaya, Saratov 410028, Russia

[*]To whom correspondence should be addressed. E-mail: (NGK) khlebtsov@ibppm.ru

ABSTRACT. Recently, modal expansion methods (MEMs) were developed to accurately predict the extinction and scattering spectra of bare and coated plasmonic particles. However, it remains uncertain whether the accuracy of analytical models is preserved when incorporating size-corrected dielectric functions. We compare numerical (COMSOL) and analytical extinction and scattering spectra for various particle shapes—rods, disks, prisms, bicones, and bipyramids—using both bulk and size-corrected dielectric functions. Size correction primarily causes broadening and a reduction in the plasmonic peaks. Still, the accuracy of analytical models is maintained for all shapes except sharp bicones, where MEM fails for both bulk and size-corrected dielectric functions. This failure arises from the strong localization of the plasmonic field near the sharp bicone tips. However, for realistic bicone models with nanometer-scale tip curvature, MEM performs well.

1. INTRODUCTION

In biomedical applications, plasmonic gold and silver nanoparticles are typically used as bioconjugates with surface-functionalized molecules or as components of hybrid nanostructures featuring a dielectric coating that embeds fluorescent, photodynamic, or targeting agents.[1] To simulate the extinction and scattering spectra of such hybrid structures, simple analytical models[2,3] based on a combination of the modal expansion method (MEM)[4] and the dipole equivalence method (DEM)[5] have recently been proposed. These models, grounded in straightforward physical principles, are essential not only for conventional simulations but also for predicting optical properties via machine learning approaches[6,7] and for in situ monitoring of synthesis processes.[8]

Comparisons with finite element method (COMSOL) calculations demonstrated the reasonable accuracy of the analytical models in capturing the main plasmonic extinction and scattering peaks for particles of various shapes.[3] However, these comparisons did not account for additional attenuation mechanisms that modify the dielectric function of metal nanoparticles compared to bulk samples. Therefore, whether the accuracy of analytical models remains when incorporating size effects on the dielectric function was an open question—one addressed in this study.

Using a well-established size-correction approach for the dielectric function[9] within the classical Drude-Lorentz framework,[10] we analyze the accuracy of analytical models for gold and silver nanoparticles of varying sizes and shapes, including rods, disks, prisms, bicones, and bipyramids. We demonstrate that size correction results in the expected broadening and reduction of plasmonic peaks, particularly for small silver nanoparticles. Nevertheless, these effects do not compromise the accuracy of the analytical models, which maintain excellent agreement with numerical COMSOL calculations.

II. THEORETICAL METHODS

A. Particle models

Figure 1 shows five NP models that are described by a characteristic size $L$, a generalized aspect ratio $AR$, and the complex metal permittivity $\varepsilon_1$. A homogeneous coating (Figure 1B) is characterized by a constant shell thickness $s$ and permittivity $\varepsilon_2$; a homogeneous host medium is characterized by a dielectric function $\varepsilon_m$ (water in this work). In the following, we used a constant permittivity, $\varepsilon_2 = 2.25$, which corresponds to a typical refractive index of 1.5 for most coatings and embedding matrices.[11]

The Supplementary Material (SM) file gives details of the particle shape geometries and explicit formulas for the core and total particle volumes (Section S1). In what follows, the core and shell characteristic sizes and aspect ratios will be designated by indices 1 and 2, respectively. For example, the metallic nanorod is characterized by its length $L_1$, diameter $d_1$, and aspect ratio $AR_1 = L_1 / d_1$, whereas the similar shell parameters are the length $L_2 = L_1 + 2s$, diameter $d_2 = d_1 + 2s$, and aspect ratio $AR_2 = L_2 / d_2$. We shall also use more compact designations for the core and shell aspect ratios $h_i \equiv AR_i, i = 1, 2$.

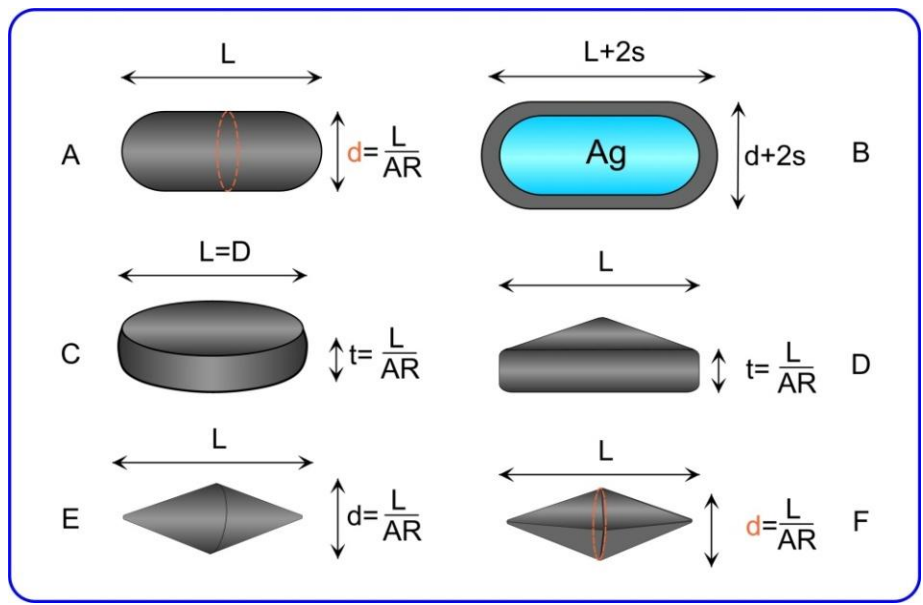

**Figure 1**. Geometrical models for five particle morphologies are characterized by the characteristic length $L$ and a generalized aspect ratio $AR$. Panel B illustrates a nanoparticle coating with a dielectric shell of constant thickness. Nanorods (NRs) have hemispherical caps at the tips (A);

Nanodisks (NDs) have semicircular edge profile (C); Nanotriangles (NTs) have smooth edges with rounding radius $(L+t)/40$ (D); Bicones (BCs) have rounded tips and circular base edge with rounding radii $d/40$ and $d/100$, respectively, (E); Bipyramids (BPs) have pentagonal smooth base and rounded tips as indicated in the Supplementary Material file (Section S1). The BP length, diameter, and aspect ratio correspond to those of the inscribed bicone.

To exclude unphysical sharpness of boundaries for prisms and bicones, we used some small smoothing parameters suggested by Yu et al.[4] and described in the Supplementary Material file. Further, the shape of pentagonal unrounded bipyramids can be conveniently characterized by the length and diameter of the inscribed bicone.[12] Moreover, the experimental as-prepared and chemically etched bipyramids have variable rounding radii that differ significantly from the previously suggested parametrization model.[4] Specifically, the tip $R_{tip}$ and base $R_{base}$ radii increase with the decrease of the rounded BP length.[12] As the BP diameter remains almost constant during chemical etching, experimental data can be approximated by simple linear relationships

$$R_{tip}(\text{nm}) = 14 - 2AR_1, \quad R_{base}(\text{nm}) = 16 - 2AR_1, \quad AR_1 \leq 6. \quad (1)$$

This parametrization implies that the maximal aspect ratio of as-prepared BPs is less than six, in agreement with the experimental data.[12]

**B. Polarizability of coated plasmonic particles**

According to the MEM+DEM method,[3] the generalized polarizability of a coated plasmonic particle is calculated by the following expansion over modes $j$:

$$\alpha_{av} = R_{2ev}^3 \sum_j q_j(h_2) \frac{\bar{\varepsilon}_j - \varepsilon_m}{3\varepsilon_m + 3L_{2mj}(\bar{\varepsilon}_j - \varepsilon_m)}, \quad (2)$$

$$L_{2mj} = \frac{1}{1 - \eta_j(h_2)} - A_j(h_2, \varepsilon_m), \quad (3)$$

$$A_j(h_2, \varepsilon_m) = a_{j2}(h_2) x_{2m}^2 + i \frac{2}{9} q_j(h_2) X_{2m}^3 + a_{j4}(h_2) x_{2m}^4, \quad (4)$$

$$x_{2m} = \frac{L_2 k_m}{2\pi}, \quad X_{2m} = R_{2ev} k_m, \quad k_m = k\sqrt{\varepsilon_m}, \quad (5)$$

where $R_{2ev}$ and $L_2$ are the equivolume sphere radius of the shell and its characteristic size, respectively, $q_j(h_2) = V_j^m(h_2)/V_2$ are the modal volume fractions, $V_j^m(h_2)$ and $V_2$ are the modal and total volumes of a coated particle, respectively. Note that in work[3], an analogue of equation (3) contains a parameter $q_{2j}$ corresponding to $q_j(h_2)$ in this work. For particles without coating, i.e., for metal cores, the above equations reduce to the following ones:

$$\alpha_{av} = R_{1ev}^3 \sum_j q_j(h_1) \frac{\varepsilon_1 - \varepsilon_m}{3\varepsilon_m + 3L_{1mj}(\varepsilon_1 - \varepsilon_m)}, \tag{2a}$$

$$L_{1mj} = \frac{1}{1 - \eta_j(h_1)} - A_j(h_1, \varepsilon_m), \tag{3a}$$

$$A_j(h_1, \varepsilon_m) = a_{j2}(h_1)x_{1m}^2 + i\frac{2}{9}q_j(h_1)X_{1m}^3 + a_{j4}(h_1)x_{1m}^4, \tag{4a}$$

$$x_{1m} = \frac{L_1 k_m}{2\pi}, \quad X_{1m} = R_{1ev}k_m, \quad k_m = k\sqrt{\varepsilon_m}, \tag{5a}$$

where $R_{1ev}$ and $L_1$ are the equivolume sphere radius of the core and its characteristic size, respectively, $q_j(h_1) = V_j^m(h_1)/V_1$ are the modal volume fractions, $V_j^m(h_1)$ is the modal volume of the core, and $V_1$ is the core volume.

The eigenvalues $\eta_j(h_1)$ and $\eta_j(h_2)$ of j-modes ($j=1$ corresponds to the dipole eigenmode) are the key parameters of MEM that depend on the particle shape only, or, in other words, on the generalized aspect ratios $h_i = AR_i$, $i=1,2$. However, they are independent of the particle size, composition, or environment, thus making the MEM approach quite universal. Explicit analytical approximations for MEM parameters are given in the Supplementary Material file (Section S1).

Looking at the MEM+DEM expansion (2), it is easy to see that it is nothing but a sum of modal polarizabilities, the form of which coincides with the polarizability of a spheroid with average permittivity $\bar{\varepsilon}_j$ and generalized depolarization factor $L_{2mj}$. In the case of a bare metal particle, the parameter $\bar{\varepsilon}_j$ is replaced by the metal dielectric function $\varepsilon_1$, and the generalized depolarization factors of the core are calculated using equation (3), where the index 2 is replaced with 1. Thus, Eq. (2) reduces to the original MEM expansion[4,13,14] but in its recast form.[3]

For coated particles, we combine MEM with DEM, retaining the formal structure of the MEM expansion (2). However, we now need to calculate the averaged modal dielectric functions $\bar{\varepsilon}_j$ for a given shell thickness $s$ and its dielectric permittivity $\varepsilon_2$. Explicit formulae for $\bar{\varepsilon}_j$ were obtained previously[3]

$$\bar{\varepsilon}_j = \varepsilon_2 \left(1 + \frac{B_{12j}}{1 - L_{22j} B_{12j}}\right), \tag{6}$$

$$B_{12j} = f_{12} \frac{q_j(h_1)}{q_j(h_2)} \frac{\varepsilon_1 - \varepsilon_2}{\varepsilon_2 + L_{12j}(\varepsilon_1 - \varepsilon_2)}, \tag{7}$$

where $f_{12} = (R_{1ev} / R_{2ev})^3$ is the core volume fraction. Explicit expressions for the effective depolarization factors and related quantities are as follows:

$$L_{i2j} = \frac{1}{1 - \eta_j(h_i)} - A_j(h_i, \varepsilon_2), \quad i = 1, 2, \tag{8}$$

$$A_j(h_i, \varepsilon_2) = a_{j2}(h_i) x_{i2}^2 + i \frac{2}{9} q_j(h_i) X_{i2}^3 + a_{j4}(h_i) x_{i2}^4, \quad i = 1, 2, \tag{9}$$

$$x_{i2} = \frac{L_i k_2}{2\pi}, \quad X_{i2} = R_{iev} k_2, \quad i = 1, 2, \tag{10}$$

where $k_2 = k\sqrt{\varepsilon_2} = (2\pi/\lambda)\sqrt{\varepsilon_2}$, $L_1$ and $L_2$ are the characteristic sizes of the core and shell, respectively; the modal volume fractions $q_j(h_i)$ depend on the aspect ratios of core ($h_1$) and shell ($h_2$), respectively. The parameters $q_j(h_i)$ obey the sum rule $\sum_j q_j(h_i) = 1$, $i = 1, 2$. In the above Eqs. (6-7), the parameters $L_{12j}$ and $L_{22j}$ correspond to the generalized depolarization factors calculated by the MEM equations (8). Note that $L_{22j}$ factors are determined by the core eigenvalues $\eta_j(h_2)$ for the shell aspect ratio $h_2$.

The first term in Eq. (4) corresponds to the shape-dependent retardation correction,[15,16] whereas the second one describes the radiation damping[17] in terms of the shape-independent equivolume shell radius parameter $X_{2m} = k_m R_{2ev}$ multiplied by the modal volume fractions. The last term in Eq. (4) accounts for small contributions $\sim x_{1m}^4$ multiplied by shape-dependent modal coefficients $a_{j4}(h_1)$.

The MEM parameters $\eta_j$, $V_j^m$, $a_{j2}$, $a_{j4}$ are calculated by minimizing the deviations of the MEM solution from benchmark numerical simulations. A remarkable advantage of MEM is that the fitting parameters for metal particles depend only on the generalized aspect ratio $h_i \equiv AR_i$ and do not depend on the particle size, composition, or environment. In summary, MEM provides straightforward and accurate analytical models to simulate extinction, scattering spectra, and other electromagnetic properties of arbitrarily shaped plasmonic NPs. Due to their analytical simplicity, MEM models can be easily implemented on any desktop PC to simulate plasmonic responses from realistic experimental models involving size and shape statistical distributions, which are complicated by random orientations.

The MEM polarizability (3) is calculated for a specific polarization of the incident light in the particle frame. The extinction and scattering cross sections averaged over random particle orientations are[18]

$$\langle C_{ext} \rangle = \pi R_2^2 \langle Q_{ext} \rangle = \frac{4\pi}{k'_m} \mathrm{Im}\left( k_m^2 \frac{\alpha_{av}^x + \alpha_{av}^y + \alpha_{av}^z}{3} \right), \quad (11)$$

$$\langle C_{sca} \rangle = \pi R_2^2 \langle Q_{sca} \rangle = \frac{8\pi}{3} |k_m|^4 \frac{|\alpha_{av}^x|^2 + |\alpha_{av}^y|^2 + |\alpha_{av}^z|^2}{3}, \quad (12)$$

where $k'_m = \mathrm{Re}(k_m) = (2\pi/\lambda)\mathrm{Re}(n_m)$, the angular brackets designate orientation averaging, and superscripts *x, y, z* stand for polarizations of the incident light along the corresponding frame axes related to the particle. Equations (11-12) are applicable for an arbitrary absorbing host, whereas for a dielectric environment, they reduce to known results.[19]

**C. Size-Dependent Correction of the Particle Dielectric Function**

Apparently, Doyle[20] was the first author to propose the following simple hypothesis. The resonance widths he measured for NaCl crystals—ranging from .05 to .5 eV—were significantly larger than the calculated value of .02 eV obtained using the bulk dielectric function. Doyle suggested that these measured resonance widths correspond to a limited average electron path length between 1.2 and 3.6 nm, due to electron collisions with the particle boundary: "Since the particle diameter is much smaller than the mean free path in bulk material, we may assume that each free path begins and ends at the surface." In other words, the macroscopic electron mean free path in a bulk sample should be replaced by the particle size. In the same year, 1958, Hampe published a study in which he measured the absorption spectra of gold nanoparticles in a silicate matrix.[21] He also observed broadening of the spectra and suggested that it could be related to the limitation of the electron mean free path. Three years later, Römer and Fragstein[22] measured, for the first time, the real and imaginary parts of the refractive index of gold nanoparticles with particle sizes of 3-4, 13-15, and 26-30 nm (this work extends their previous study[23]). They found significant deviations from the bulk optical constants for the smallest 3-4-nm particles. The authors explained those deviations by the limited mean free path of electrons, in agreement with earlier suggestions by Doyle[20] and Hampe.[21] In its completed form, the classical concept of changes in optical constants due to the limitation of the electron mean free path was developed in 1969 by Kreibig and Fragstein.[24] In particular, they suggested the following size-corrected expression for the dumping constant $\gamma(a)$ in a spherical particle of radius $a$

$$\gamma(a) = \gamma_b + \gamma_s = \gamma_b + A_s \frac{v_F}{a}, \qquad (13)$$

where $\gamma_b$ is the dumping constant of bulk metal, and $v_F$ is Fermi velocity, and the scattering constant $A_s$ depends on the electron surface scattering model. Starting with these early works, the classical concept of limiting the mean free path in nanoparticles has been successfully applied in numerous studies[25-28] (see also references in the reviews[29,30]).

Kawabata and Kubo[31] pointed out that interpreting the size effect as a result of electron scattering on the surface of nanoparticles is incorrect. From a quantum-mechanical (QM) point of view, the size effect results in the appearance of surface discrete levels, and dipole transitions between them determine the increase in the imaginary part of the particle dielectric function. Surprisingly enough, the QM calculation by Kawabata and Kubo and other authors[32-34] essentially led to the same formula (13), with a coefficient of the order of unity.

In general, three possible QM size effects should appear in metal nanoparticles[34]: a blue shift of the localized plasmon resonance (LPR) wavelength, LPR broadening, and a fine structure of the absorption spectrum. The third effect has not been observed experimentally due to the size distribution in real samples. As for the LPR shift and broadening, these effects have been observed experimentally,[26,28,29] but they can be satisfactorily explained in terms of the classical size-limited mean free path of electrons. Therefore, here we adopt the following classical procedure to correct the tabulated dielectric function of bulk metals $\varepsilon_b(\omega)$:

$$\varepsilon(\omega, a) = \varepsilon_b(\omega) + \Delta\varepsilon(\omega, a), \qquad (14)$$

the correction $\Delta\varepsilon(\omega,a)$ accounts for the contribution of size-dependent electron scattering to the Drude part of the dielectric function

$$\Delta\varepsilon(\omega,a) = \frac{\omega_p^2}{\omega(\omega+i\gamma_b)} - \frac{\omega_p^2}{\omega(\omega+i\gamma_a)}, \tag{15}$$

where $\omega_p$ is the plasma frequency and the particle dumping parameter $\gamma_a$ consists of three contributions

$$\gamma(a) = \gamma_b + \gamma_s + \gamma_{CID}. \tag{16}$$

Bulk dumping $\gamma_b$ is due to scattering with phonons, impurities, electrons, etc., and it is assumed to be size-independent and the same as in the bulk metals. The second term $\gamma_s$ describes the surface scattering of electrons, while the third term $\gamma_{CID}$ describes an additional chemical interface damping (CID) caused by changing the nanoparticle's chemical interface.[35,36] Note that we did not include the radiation dumping term $\gamma_{rad}$[37,38] in Eq. (16) because we use the full-wave FEM solution with COMSOL, where the radiation dumping is included automatically. The CID dumping should scale inversely with particle size as the number of $s$-electrons interacting with the surface adsorbate is proportional to the particle volume $V$, whereas the number of adsorbate molecules is proportional to the particle surface area $S$. Then, the last two terms in Eq. (16) can be combined into one relation

$$\gamma_s = \left(A_s^{surf} + A_s^{CID}\right)\frac{v_F}{l_s} = A_s \frac{v_F}{l_s}, \tag{17}$$

where $l_s$ is calculated by the Coronado and Schatz[39] recipe

$$l_s = \frac{4V}{S}. \tag{18}$$

Table 1 summarizes the effective scattering lengths for selected particle shapes (for details, see Supplementary Material file, Section S2)

**Table 1. Effective scattering length $l_s$ for particles of various shapes.**

| Particle shape | Scattering length, $l_s$ | Comments |
|---|---|---|
| Rods with semispherical | $d\left(1-\dfrac{1}{3h}\right)$ | The total length, diameter, and aspect ratio are $L$, $d$, and $h=L/d$, respectively. |
| ends | | |
| Disks with circular edges | $t\dfrac{2}{3}\dfrac{1+\dfrac{3}{4}(h-1)(\pi+2h-2)}{1+\dfrac{h-1}{2}(\pi+h-1)}$ | The total diameter, thickness, and aspect ratio are $D$, $t$, and $h=D/t$, respectively. |
| Bicones and pentagonal bipyramids | $\dfrac{2}{3}d\dfrac{h}{\sqrt{1+h^2}}$ | For bipyramids, the total length $L$, diameter $d$, and aspect ratio $h=L/d$ are defined for an inscribed bicone. |

Under conditions $\gamma_s^2/\omega^2 \ll 1$, $\gamma_b^2/\omega^2 \ll 1$, Eq. (15) reduces to:

$$\Delta\varepsilon(\omega,l_s) = A_s \frac{\omega_p^2}{\omega^3}\frac{v_F}{l_s}(B_s+i). \tag{19}$$

$$B_s = \frac{(2\gamma_b + A_s v_F/l_s)}{\omega}, \tag{20}$$

According to the literature data for bulk gold,[40] the average values of plasma frequency and damping constant are $\hbar\omega_p = 9.0 \pm 0.38$ eV and $\gamma_b = (1.02 \pm 0.17)\times 10^{14}$ s$^{-1}$, respectively. Practically, one can set $B_s = 0$ and use a pure imaginary correction (19). Similarly, for silver, we employed an imaginary correction (19) as suggested by Kreibig.[26] In this case, the bulk parameter $\gamma_b$ is not used for correction, and the only input parameter is the scattering constant $A_s$. There is a notable spread in published theoretical QM and experimental $A_s$ values as reviewed by Kreibig and Genzel.[29] Recent data on extinction of Au nanorod colloids,[41] absorption of single Au nanospheres,[42] and scattering by single Au nanorods,[38,43,44] suggest the best fitting value $A_s \approx 1/3$. This estimation should be considered a lower limit for CTAB (cetyltrimethylammonium bromide) coated Au nanorods (for details, see Ref.[40]). For silver particles, we used $A_s = 1$, which aligns closely with most experimental data for hydrosols and Ag particles in different matrices.[35] Note that the values $A_s = 2.5$ for silver and $A_s = 2$ for gold used by Kondorskiy and Mekshun[45] in MEM simulations of triangle nanoprisms appear to be greatly overestimated.

The dielectric functions of gold and silver were calculated as described previously by tabulated data[46] and by a spline described in the Supporting Information file of Ref.[2], respectively.

## II. RESULTS AND DISCUSSION

### A. Gold and silver rods

For all the calculations discussed below, the diameter of the nanorods was 15 nm, which is a typical value for most particles used in various biomedical applications.[47] Considering that for biomedical applications, the plasmonic resonance should be tuned within the first transparency window of biological tissues (700-950 nm),[48] we limited our discussion here to aspect ratios of 4 and 5, although overall, for aspect ratios of 2-6, the results do not differ significantly. Before conducting large-scale computations for coated rods, the accuracy of the COMSOL calculations was verified through benchmarking simulations for bare particles with TMM codes.[18] For all cases tested, we found excellent agreement between both methods except for some slight differences in multipolar peaks of silver nanorods.[3]

Figure 2 shows a comparison between numerical and analytical extinction spectra for randomly oriented gold and silver NRs, both uncoated and with a 30 nm (A–D) and 60 nm (E, F) shell. In all cases, volume-based and size-corrected optical constants of the metals were used. From the results of these calculations, several conclusions can be drawn.

As expected, size-dependent damping results in a reduction and broadening of the plasmon peaks, particularly for silver NRs. Secondly, the analytical model demonstrates good agreement with numerical calculations for metallic NRs when using both volume-based and size-corrected optical constants. Thirdly, the analytical model for metallic nanorods, combined with the DEM, provides high accuracy in describing the extinction spectra of coated NRs, effectively accounting for size-dependent damping. Fourthly, the MEM+DEM model can be applied to NRs with a thick 60-nm shell, even when the volume fraction of the metal core is as low as 0.004. In summary, based on our large-scale calculations, the MEM+DEM analytical model is valid not only for extinction spectra but also for scattering spectra (see Figure S3 in the Supplementary Material file). It works well for particles with coatings up to 100 nm thick,[3] aspect ratios ranging from 1 to 6, and maximum sizes (lengths) around 200 nm. The major limitation is that the MEM+DEM method is primarily effective near the main plasmon peak and does not describe minor multipolar peaks.

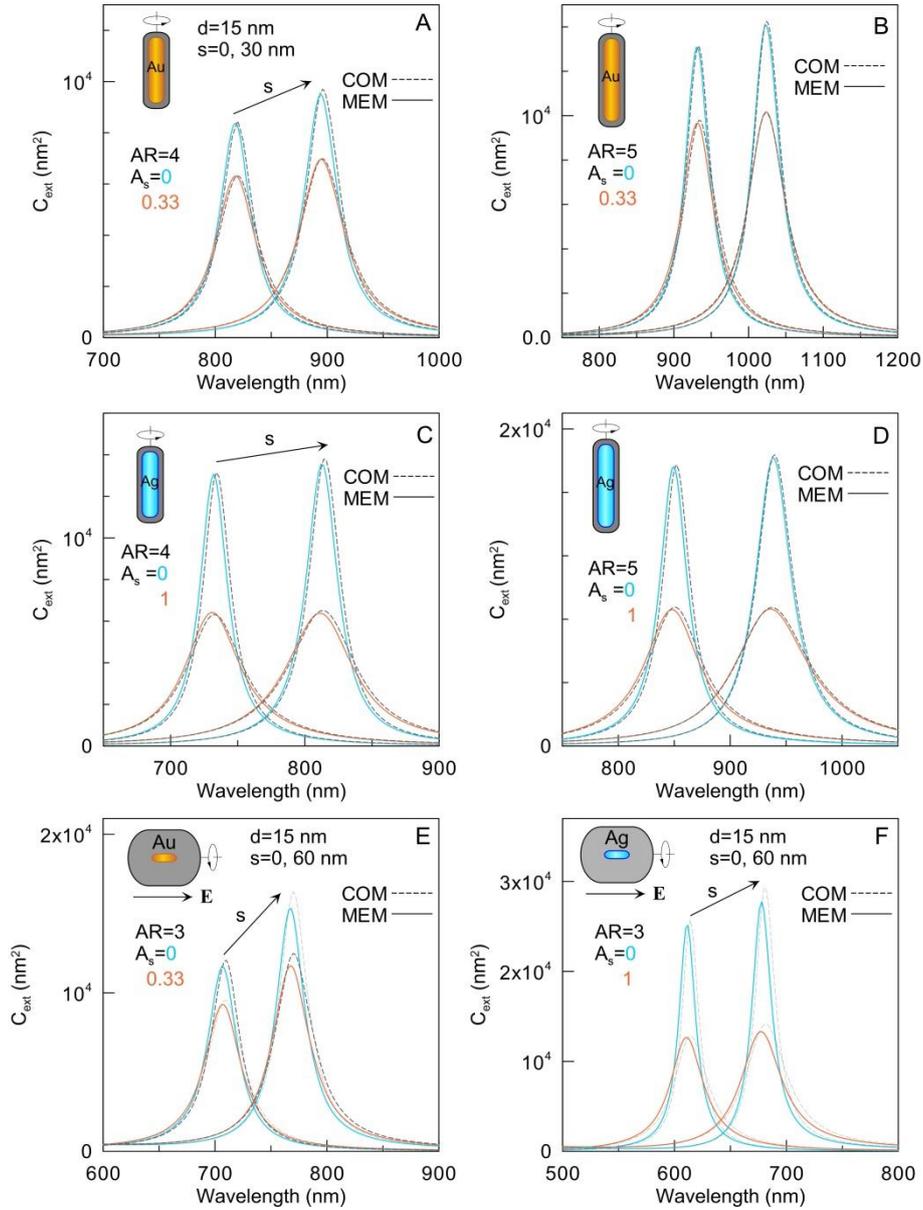

**Figure 2**. Extinction spectra of gold (A, B, E) and silver (C, D, F) bare ($s = 0$) and coated ($s = 30$ (A–D) and 60 nm (E, F)) NRs with aspect ratio 4 (A, C), 5 (B, D), and 3 (E, F), and a fixed diameter of 15 nm. Calculations by COMSOL (dashed lines) and MEM (solid lines) for bulk ($A_s = 0$, blue) and size-corrected ($A_s = 0.33$ for gold and $A_s = 1$ for silver, orange) optical constants. The electron scattering constants are indicated in the panels. Spectra in panels A–D are calculated for randomly oriented rods, whereas panels E and F show longitudinal excitation spectra.

## B. Gold and silver disks

Figure 3 shows extinction spectra for bare and coated gold and silver NDs under in-plane excitation. For the smallest dimensions ($10 \times 60$ nm), there is excellent agreement between the numerical and analytical spectra calculated using both bulk and size-corrected optical constants. The analytical model provides accurate predictions for both the reduced extinction peak and spectral broadening. When the disk size increases to $20 \times 120$ nm, the spectra for size-corrected and bulk optical constants become nearly identical due to negligible electron scattering effects. However, the accuracy of the analytical (MEM+DEM) model degrades slightly: the broadening is overestimated compared to numerical results, and the peak position shows a small redshift.

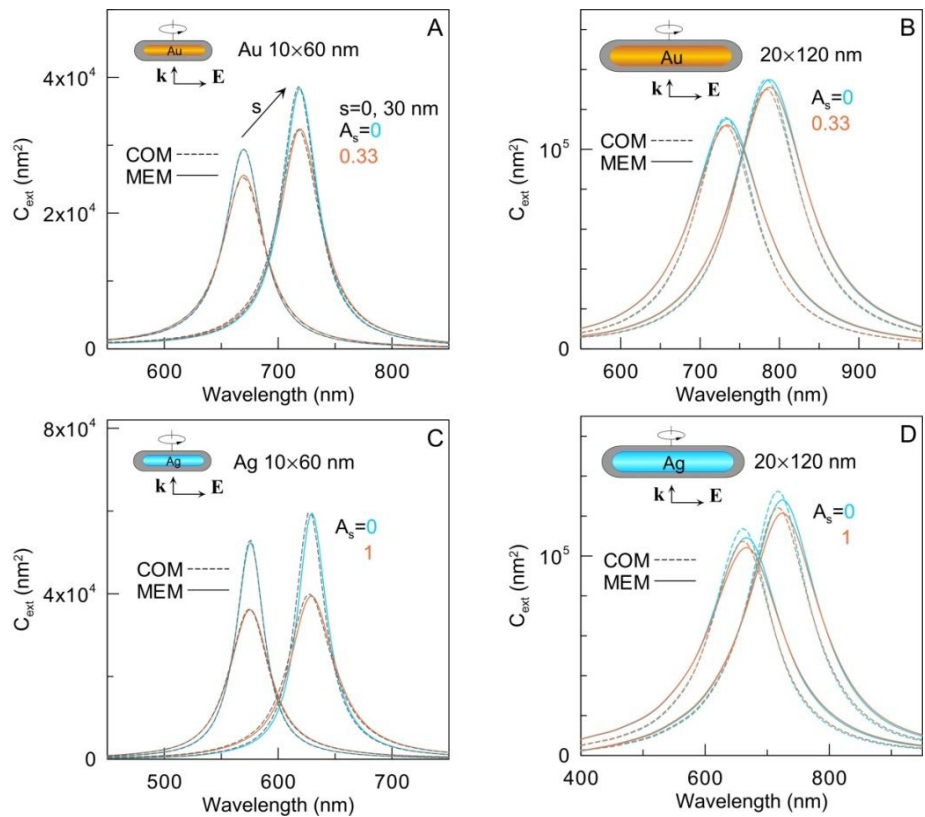

**Figure 3**. Extinction spectra of gold (A, B) and silver (C, D) bare and coated ($s = 30$ nm) NDs with dimensions of $10 \times 60$ nm (A, C) and $20 \times 120$ nm (B, D). Calculations by COMSOL (dashed lines) and MEM (solid lines) at in-plane excitation for bulk ($A_s = 0$, blue) and size-corrected optical constants ($A_s = 0.33$ for gold and $A_s = 1$ for silver, orange).

Similar properties are observed in the scattering spectra (Figure S4 in the Supplementary Material file). In summary, the analytical model can simulate the extinction and scattering spectra of gold and silver disk-shaped particles, provided their dimensions do not exceed $20 \times 120$ nm and the coating thickness is less than 60 nm. Size correction of optical constants is only necessary for thin disks with thicknesses of about 10 nm.

## C. Gold and silver nanotriangles

Extinction spectra of thin 10-nm gold and silver NTs (Figure 4 A, B) demonstrate notable broadening due to additional size-correction dumping, especially for silver particles.

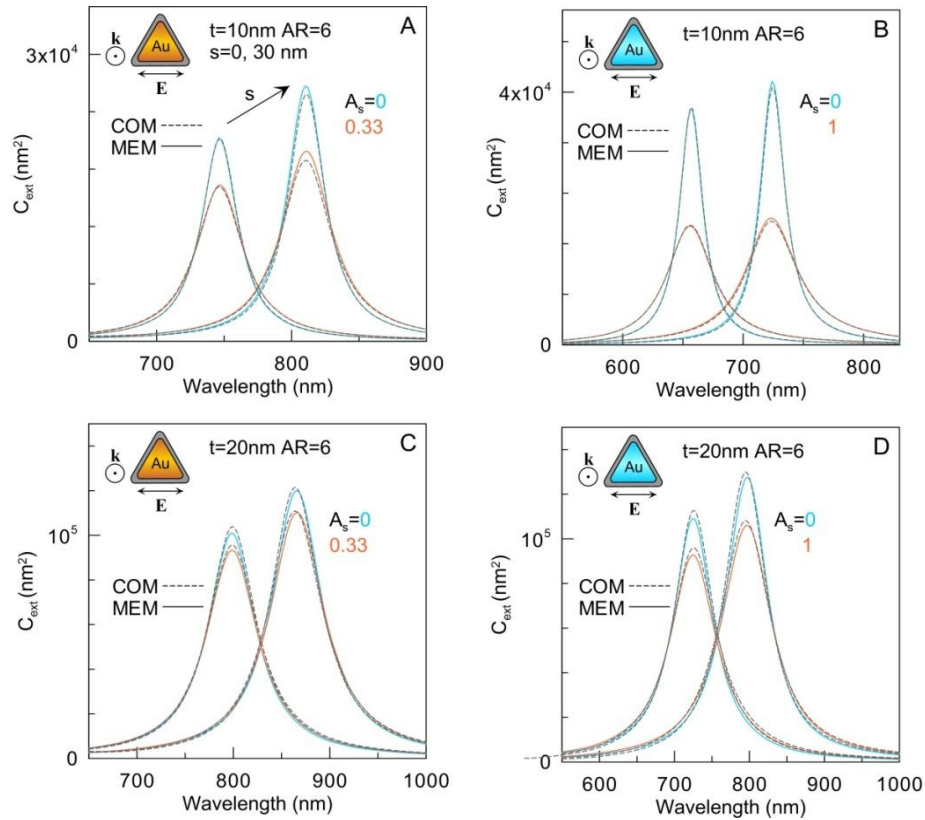

**Figure 4**. Extinction spectra of gold (A, C) and silver (B, D) bare and coated ($s = 30$ nm) NTs with dimensions of $10 \times 60$ nm (A, B) and $20 \times 120$ nm (C, D). Calculations by COMSOL (dashed lines) and MEM (solid lines) under in-plane excitation for bulk ($A_s = 0$, blue) and size-corrected optical constants ($A_s = 0.33$ for gold and $A_s = 1$ for silver, orange).

Similar to rods and disks, the MEM+DEM analytical model performs nearly perfectly in predicting peak positions and broadening for both bare and coated NTs. The same behavior is observed in the scattering spectra (Figure S5 in the Supplementary Material file). For larger NTs measuring $20 \times 120$ nm, size-correction effects become negligible from a practical perspective, especially for gold. However, unlike NDs, the analytical MEM+DEM model for NTs still shows close agreement with COMSOL-simulated spectra. This result is somewhat unexpected, as nanodisks have a more regular and smoother shape compared to triangular nanoprisms. Therefore, the developed analytical model can be applied to NTs with larger core sizes and shell thicknesses than comparable oblate nanodisks.

**D. Gold and silver bicones**

For bare gold and silver bicones (BCs), our analytical MEM model (Supplementary Material file, Section S1.4) accurately reproduces numerical extinction spectra near the main LPR peaks.[3] However, significant differences arise between numerical and analytical MEM+DEM spectra calculated for coated BCs. Specifically, numerical spectra exhibit a strong LPR shift even with a slight 1-nm coating, in stark contrast to MEM+DEM simulations, which predict a much weaker shift. We will discuss this discrepancy and its possible physical origins later in Section F. Here, we consider a factorized version of the MEM+DEM approach,[3] in which the MEM parameter $\eta_1(h_i)$ and the normalized modal volumes $V_1^m(h_i)/V_i$ for the first dipole mode can be represented in a factorized form.

$$\eta_1(h_i) = f_{\eta h}(h_i) f_{\eta s}(s), \tag{21}$$

$$V_1^m(h_i)/V_i = f_{Vh}(h_i) f_{Vs}(s), \tag{22}$$

where the first factors correspond to the MEM parameters calculated for the aspect ratio of the core ($h_1$) or shell ($h_2$) according to the formulas specified in the Supplementary Material file. Note that the modal volumes are normalized to the bare or coated bicone volumes $V_i$. The second factor in equations (21-22) accounts for the specific shell effects and satisfies the apparent conditions $f_{\eta s}(0) = 1$ and $f_{Vs}(0) = 1$. Functions $f_{\eta s}(s)$ and $f_{Vs}(s)$ account for the shell effects and are independent of the particle size, shape, and metal composition. In this sense, the factorized correction (21-22) is universal, eliminating the need for additional fitting of analytical spectra for a particular particle shape and shell thickness.

Figure 5 illustrates the accuracy of the factorized MEM+DEM analytical model when applied to coated BCs. Consistent with previous results, the analytical extinction spectra show excellent agreement with COMSOL simulations performed using both bulk and size-corrected optical constants. Similar results were also obtained for the light scattering spectra of coated bicones with these optical constants (see Figure S6 in the Supplementary Material file). In these simulations, we employed a geometric model with a fixed length of 100 nm and a variable aspect ratio to adjust the LPR peak across the UV-vis region. For small aspect ratios below 2, size-correction effects are minimal and can be neglected in practical comparisons of experimental and theoretical LPR peaks. However, when the aspect ratio of the BC increases to 4, its diameter becomes small enough to cause size-dependent reduction and broadening of the LPR peak, which can no longer be ignored. Similar to particles of other shapes, size effects are more pronounced for silver compared to gold. The agreement between the analytical and numerical spectra in panels B and D is nearly perfect, owing to the smaller particle diameter and volume (by a factor of 4). It should be noted that for strongly elongated silver bicones (panel D), the extinction LPR peaks decrease with increasing shell coating thickness. The physical origins of this effect have been discussed in our previous study.[3]

Additionally, we tested the factorized model with a fixed BC diameter of 30 nm and a variable aspect ratio. This model successfully fits the experimental data for chemically etched bipyramids.[12] Once again, we achieved good agreement between the analytical and numerical extinction and scattering spectra for both bulk and size-corrected optical constants of gold and silver cores. To summarize, the factorized MEM+DEM analytical models can be applied to simulate the extinction and scattering spectra of coated gold and silver bicones with sharp tips.

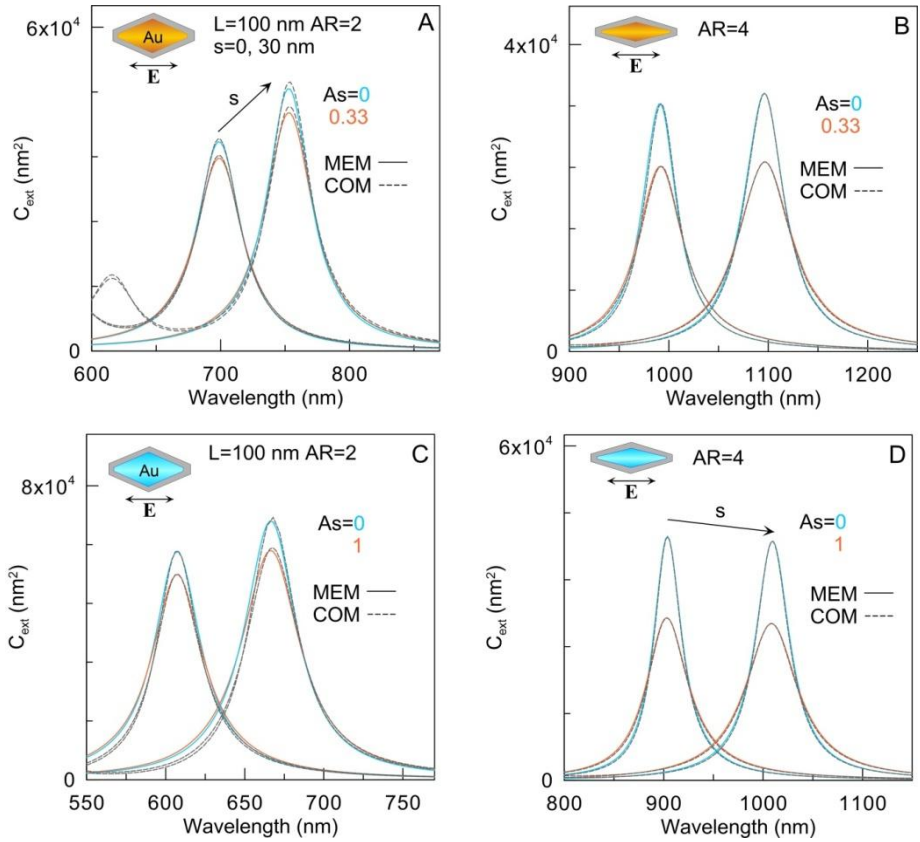

**Figure 5**. Extinction spectra of gold (A, B) and silver (C, D) bare and coated ($s = 30$ nm) bicones with a fixed length of 100 nm and aspect ratios of 2 (A, C) and 4 (B, D). Calculations by COMSOL (dashed lines) and MEM (solid lines) under longitudinal excitation for bulk ($A_s = 0$, blue) and size-corrected optical constants ($A_s = 0.33$ for gold and $A_s = 1$ for silver, orange).

## D. Gold and silver bipyramids

Currently, gold nanorods remain the most well-known and widely used plasmonic particles.[47] In contrast, pentagonal gold and silver bipyramids (BPs) have been much less studied, despite exhibiting similar LPR tunability through aspect ratio variation. Structurally, gold BPs differ significantly from nanorods, featuring five equally spaced twinning planes aligned along the symmetry axis and sharp tips, whereas nanorods are single crystals with semispherical ends. These morphological differences lead to distinct local plasmonic properties, including higher environmental sensitivity[49] and superior quality factors in absorption and scattering peaks.[50] Despite these advantages, gold BPs remain surprisingly underutilized, primarily due to synthesis challenges[51-54] – even two decades after their initial report.[55] Furthermore, modeling their optical properties (particularly with molecular coatings) has historically required numerical methods, as simple, accurate analytical models were unavailable until recently. This work partially addresses this critical gap.

The shape morphology of pentagonal bipyramids (BPs) is complex to correlate with transmission electron microscopy (TEM) images. Recently, Montaño-Priede et al.[12] introduced a geometrical inversion approach using an inscribed rounded bicone (BC) model that reproduces the main TEM-based BP shape features: length, diameter, and tip curvature. This model has been successfully applied[3,12] to simulate extinction spectra of experimental samples obtained by chemical etching. A key feature of this model, derived from TEM images, is that tip and base radii increase as total particle length decreases. Our parametrization simulates this property through Eq. (1). We emphasize the fundamental difference between our model and that used in the original MEM paper by Yu et al.[4], where the BP tip and base rounding radii were significantly smaller than those of experimental particles.[12] For complete details on gold bipyramid geometrical characterization, we refer readers to our previous work.[3]

Figure 6 shows extinction spectra of coated bipyramids (BPs) for three aspect ratios (3, 4, and 5), positioning the LPR peaks within the critical biological window (650–1000 nm). Using a fixed diameter of 30 nm (typical experimental value) with either bare particles or 30-nm coatings, we demonstrate size-correction effects. An excellent agreement exists between the analytical and numerical extinction spectra for all plasmonic peaks, as also observed in the scattering spectra (Figure S7 in the Supplementary Material file). For gold BPs with the lowest aspect ratio, size-correction broadening is minimal, while being most pronounced in elongated silver BPs. Notably, coated gold and silver BPs with a maximum aspect ratio of 5 exhibit significant red shifts (~100 nm) in peak positions.

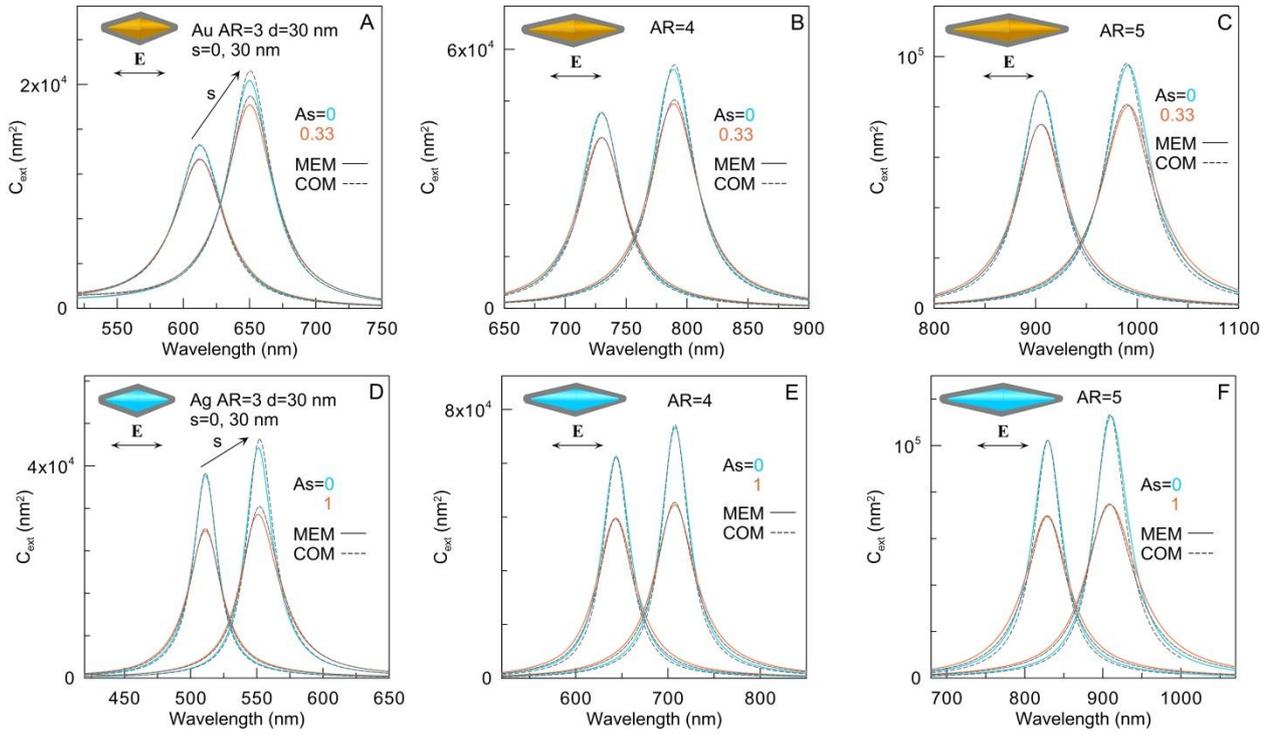

**Figure 6**. Extinction spectra of gold (A–C) and silver (D–F) bare and coated (s=30 nm) bipyramids with a fixed diameter of 30 nm, aspect ratios of 3 (A, D), 4 (B, E) and 5 (C, F), and rounding radii of tips and bases are determined by Eqs. (S30-S31). Calculations by COMSOL (dashed lines) and MEM (solid lines) under longitudinal excitation for bulk ($A_s = 0$, blue) and size-corrected optical constants ($A_s = 0.33$ for gold and $A_s = 1$ for silver, orange).

**E. Why does the MEM+DEM method fail for gold nanorods with sharp tips?**

Figures 7A and B compare numerical (COMSOL) and analytical (MEM) spectra for coated gold bicones with small rounding radii $d/40$, following the Yu et al. model.[4] Unexpectedly, we found a strong influence of the shell on spectral LPR shifts even at minimal thicknesses (1-3 nm). While MEM theory accurately describes numerical spectra for bare metallic nanoparticles, coated particles exhibit larger discrepancies in both LPR positions and amplitudes (see also Figure S8 in the Supplementary Material file). One might attribute these discrepancies to inaccuracies in modeling retardation effects within the fundamentally electrostatic MEM theory. However, this hypothesis is refuted by calculations for significantly smaller particles (20 nm total core length; Figures 7C and D). Here, coated particles exhibit even stronger differences between numerical and analytical spectra, despite near-identical spectra for bare metallic particles.

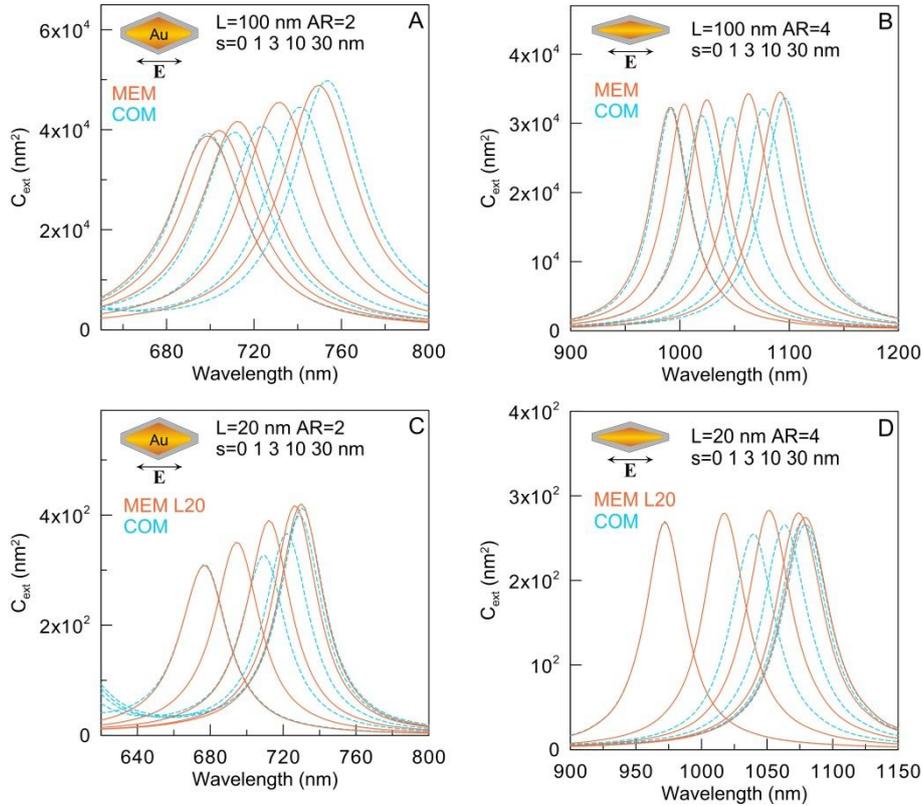

**Figure 7**. Extinction spectra of bare ($s=0$) and coated ($s=1, 3, 10,$ and $30$ nm) gold bicones with fixed lengths of 100 nm (A, B) and 20 nm (C, D), and aspect ratios of 2 (A, C) and 4 (B, D). Tip rounding radius equals $d/40$ nm. Calculations by COMSOL (blue dashed lines) and MEM (orange solid lines) under longitudinal excitation using bulk optical constants.

In Figure 8, we plot LPR peak positions as functions of shell thickness and volume fraction $g = V_{shell}/(V_{core}+V_{shell})$. For small spherical particles, the electrostatic dipole approximation[56] predicts a linear dependence between LPR position and shell volume fraction. Although 20-nm-long bicones appear to be well within the dipole approximation regime, neither numerical simulations nor analytical theory confirms the expected linear behavior LPR(g). Moreover, MEM results differ significantly from COMSOL counterparts.

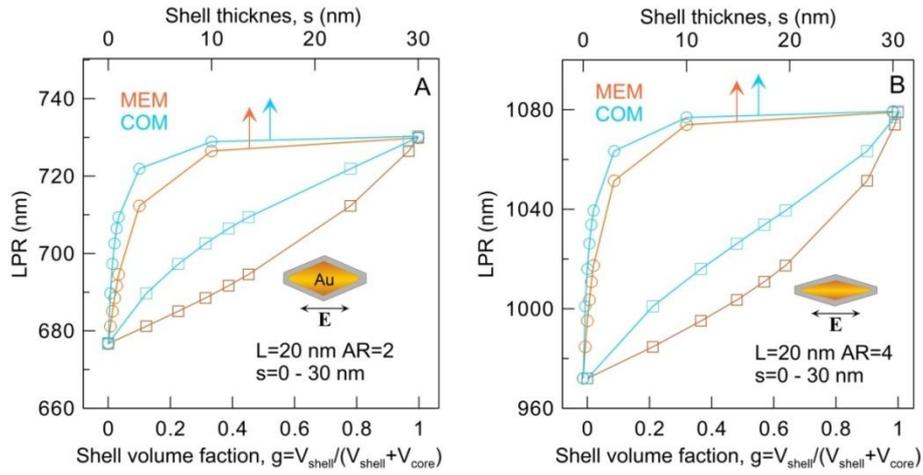

**Figure 8**. Plasmon resonance positions as functions of shell thickness (top axes) and volume fraction $g$ (bottom axes), calculated using MEM (orange curve) and COMSOL (black crosses) for bicones with total length $L = 20$ nm and aspect ratios $AR = 2$ (A) and 4 (B). The tip rounding radius is $d/40$ nm.

To explain the observed discrepancies between MEM and numerical spectra, we hypothesize that the physical reason may be related to the strongly inhomogeneous distribution of the local field around coated bicone tips. Figure 9A shows the normalized near-field distribution $\ln(E/E_{max})$ with three isolines corresponding to a decrease in the field amplitude by $e^{-n}$ with $n = 1, 2,$ and 3. This picture demonstrates strong localization of the excited near field within 0.2 nm distance from the bicone surface, where the field amplitude decreases by $e^{-1.5}$. Near the very thin 1-nm dielectric shell, the field amplitude decreases by 20 times. To quantify the local field distribution, we introduce the "local field volume fraction," defined as the ratio of the volume between the particle surface and the yellow field isoline (-1) in Figure 9B, where the field amplitude decreases by $e^{-1}$. For comparison, we evaluate the same quantity for a nanorod tip, as this model shows excellent agreement between numerical and analytical spectra for identical metallic core and shell dimensions (Figure 9C). From Figure 9A, we estimate the localization length for the bicone to be $l_{BC} \approx 0.2$ nm, while for a gold nanorod of the same length and diameter (Figure S2 in the Supplementary Material file), it is an order of magnitude larger ($l_{NR} \approx 2$ nm). A simple geometric analysis of the local field volume fractions for the bicone and the reference nanorod (Supplementary Material, Section S3) yields $\varphi_l^{BC}/\varphi_l^{NR} = 0.75 \times 10^{-2} l_{BC}/l_{NR}$. Considering the ratio $l_{BC}/l_{NR} \sim 0.1$ (Supplementary Material, Section S3), we obtain a final estimate of $\varphi_l^{BC}/\varphi_l^{NR} \sim 10^{-3}$. In summary, the plasmonic field around the bicone tip is three orders of magnitude more localized than around the nanorod, explaining the significant deviations of MEM spectra for coated particles from numerical spectra.

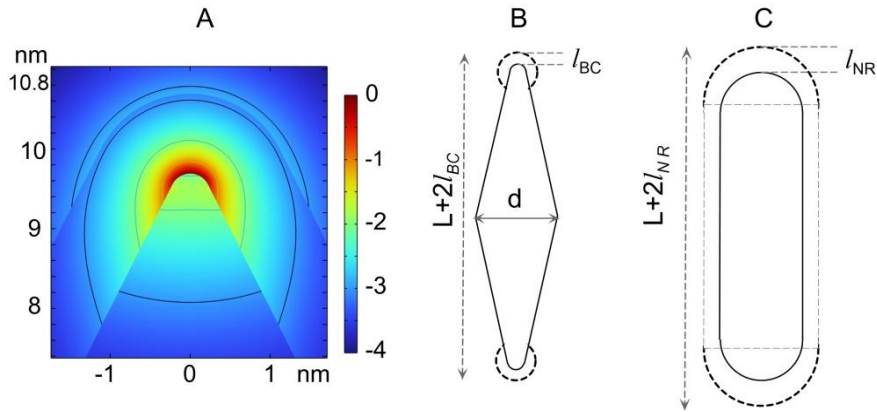

**Figure 9.** (A) Local field distribution $\ln(E/E_{max})$ around the bicone tip, where $E_{max}$ is the field magnitude at the tip surface at the resonance wavelength. Lines show where the field amplitude decreases by $e^{-n}$ for $n = 1$, 2, and 3. The bicone has a length of 20 nm and a diameter of 10 nm ($AR = L/d = 2$). Blue lines in the upper section show field isolines in the 1-nm dielectric shell and surrounding water. (B, C) Schematics for approximate evaluation of local field volume fraction for (B) bicone and (C) nanorod of identical dimensions.

Another remarkable feature of plasmonic bicones, compared to nanorods, is their ability to support multipolar excitations even at small sizes. Figure 10 shows the extinction and scattering spectra of 20-nm gold bicones with a tip rounding radius of $d/40$. Although the overall particle size is well within the dipolar Rayleigh regime, the spectra reveal two notable multipole peaks: a quadrupole excitation near 576 nm for an aspect ratio of 2 and near 716 nm for an aspect ratio of 4. These multipole resonances remain visible despite the strong dipolar scattering background from the 30-nm dielectric shell (Figures 10C and D). This is particularly surprising since no quadrupoles are observed in either 20-nm gold spheres or 20-nm rods (see inset in panel C). The effect can be attributed to both lower particle symmetry (compared to spheres) and sharper tips (compared to nanorods). Larger bicones (~100 nm in length) exhibit more pronounced, well-known multipolar features in their extinction and scattering spectra.[40]

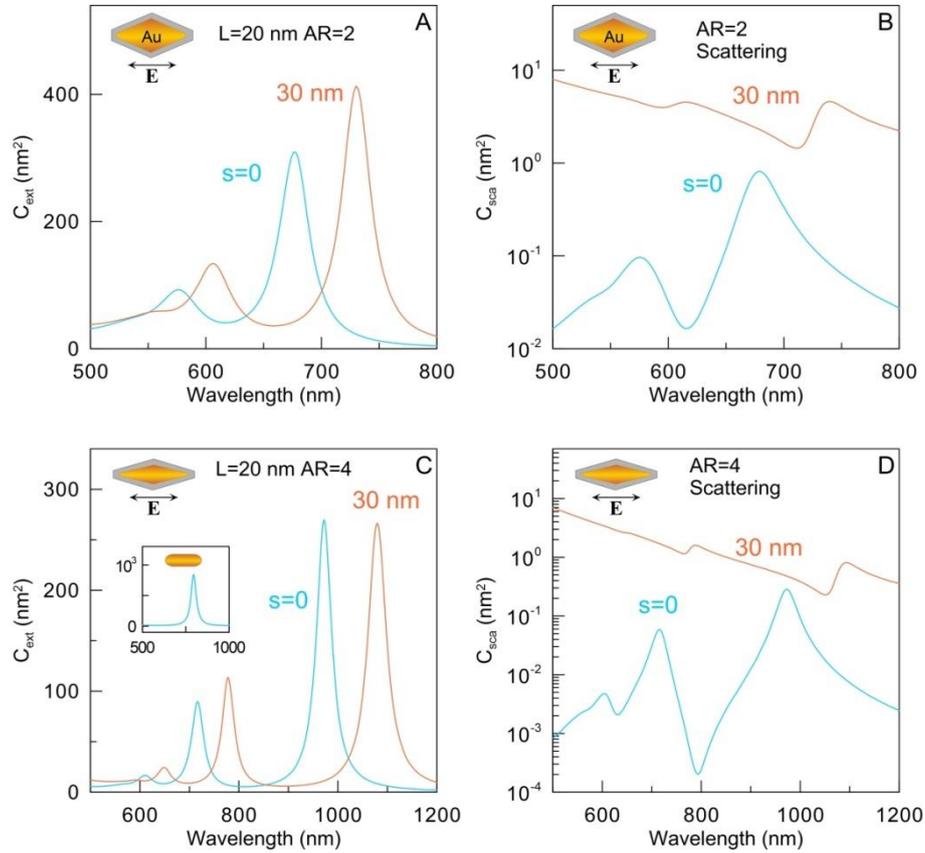

**Figure 10**. Extinction (A, C) and scattering (B, D) spectra of gold bicones with 20-nm length, aspect ratios of 2 (A, B) and 4 (C, D), and tip rounding radius of $d/40$. Numerical calculations for bare ($s=0$, blue) and coated particles ($s=30$ nm, orange). The inset in panel C shows the extinction spectrum of gold nanorods with identical length and diameter. Note the significant Rayleigh scattering from the 30-nm dielectric shell in panels C and D, where the logarithmic axes emphasize weak multipolar scattering peaks.

To provide additional evidence supporting our hypothesis, consider a bicone model derived from the experimental data of Ref. [12] In this model, the tip curvature radius can be approximated by the following relation (Supplementary Material, Eq. S27): $R_{tip} = -5AR_r + 23$, $AR_r \leq 4.5$, where the aspect ratio $AR_r$ is defined as the ratio of the actual rounded length to the particle diameter. For experimental aspect ratios ranging from 2 to 4, the tip radius varies between 13 and 3 nm—an order of magnitude larger than the tip radii predicted by the Yu et al.[4] model. For such nanometer-scale tip curvatures, the MEM+DEM model exhibits excellent agreement with numerical calculations as illustrated in Figure 11, without any factorization (see also Figure S9 in the Supplementary Material file).

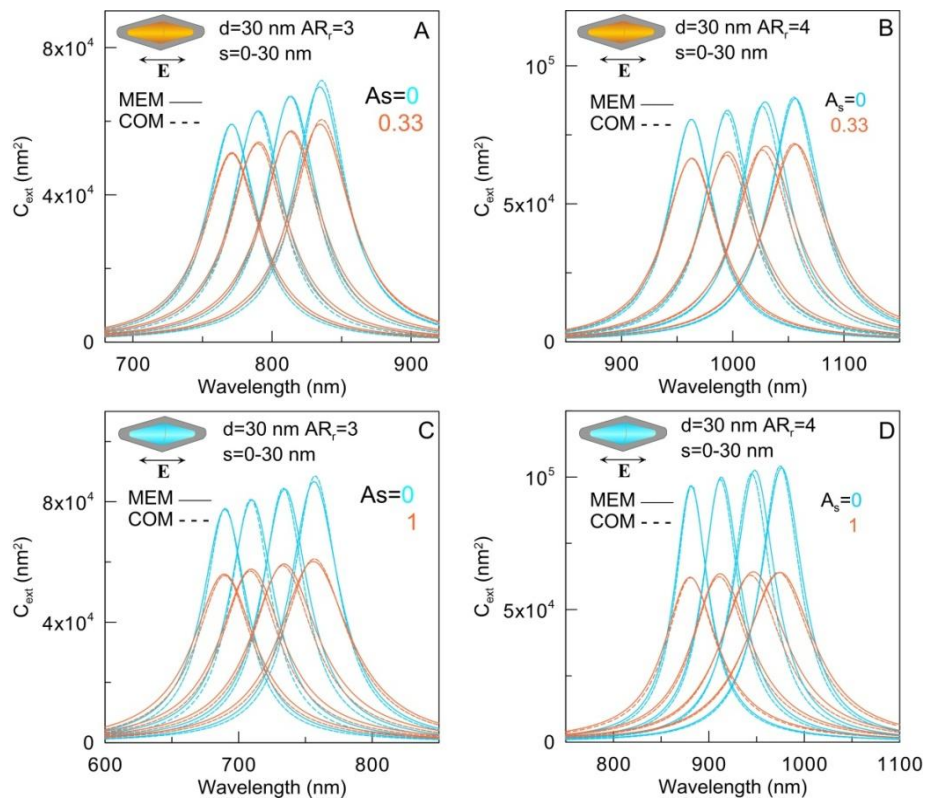

**Figure 11**. Extinction spectra of gold (A, B) and silver (C, D) bare ($s=0$) and coated ($s=3$, 10, 30 nm) bicones with a fixed diameter of 30 nm, aspect ratios of 3 (A, C) and 4 (B, D), and the tip rounding radius as given by Eq. (S27). Calculations by COMSOL (dashed lines) and MEM (solid lines) under longitudinal excitation for bulk ($A_s=0$, blue) and size-corrected optical constants ($A_s=0.33$ for gold and $A_s=1$ for silver, orange).

To summarize, the MEM+DEM model does not work for coated bicones because the usual transition to the dipolar regime (with decreasing size) is violated, and strong localized near-fields are generated within a 0.2 nm region. This explains the significant plasmonic shift observed even with a thin 1-nm shell, which cannot be predicted by the electrostatic MEM+DEM model. However, for realistic bicone tip curvatures, the MEM+DEM method works almost perfectly.

**CONCLUSION**

The question posed in the introduction—regarding the applicability of the MEM+DEM method for describing the optical properties of coated particles with size-dependent optical constants—has been answered positively. We have demonstrated that the MEM+DEM approach accurately reproduces numerical extinction and scattering spectra for various coated plasmonic particles, using either bulk or size-corrected optical constants of gold or silver. The only exception is bicones or bipyramids with sharp tips; for such particles, the factorized version of MEM+DEM must be used, reducing its universality. We have analyzed the possible physical reasons behind these unusual plasmonic properties. Surprisingly, discrepancies between numerical and analytical spectra persist even when the overall particle size is well within the dipolar Rayleigh regime (below 20 nm), ruling out retardation effects as an explanation. Notably, small 20-nm-long bicones still exhibit multipolar excitations, unlike nanorods of the same length. The most likely explanation lies in the strong localization of plasmonic fields near sharp tips. Nevertheless, the MEM+DEM approach works well when tip curvature radii increase to several nanometers. In a sense, the plasmonic responses of small bicones show some analogy to the electromagnetic peculiarities of small plasmonic cubes.[13,57]

**Supplementary Material**. Section S1: MEM parameters for selected particle shapes; Section S2. Effective scattering length of electrons. Figure S1. Section S3. Volume fraction of the local field. Figure S2. Section S3: Additional scattering spectra of bare and coated gold and silver nanorods, nanodisks, nanotriangles, bicones, and bipyramids; Figures S3-S9.

**Acknowledgments**

This research was supported by the Russian Science Foundation (project no. 24-22-00017).

## AUTHOR DECLARATIONS

**Conflict of Interest**

The authors have no conflicts to disclose.

**Author Contributions**

Nikolai Khlebtsov: Conceptualization; Formal analysis; Investigation; Writing; Supervision

Sergey Zarkov: COMSOL simulations; Data analysis

## DATA AVAILABILITY

The data supporting the findings of this study are available within the article and its supplementary materials. The T-matrix codes are freely available at https://github.com/khlebtsov/T- matrix- abs- host.

**References**

[1] N. Hlapisi, S. P. Songca, and P. A. Ajibade, "Capped plasmonic gold and silver nanoparticles with porphyrins for potential use as anticancer agents—a review," Pharmaceutic 16, 1268 (2024).

[2] N. G. Khlebtsov and S. V. Zarkov, "Analytical modeling of coated plasmonic particles," J. Phys. Chem. C 128, 15029−15040 (2024).

[3] N. G. Khlebtsov and S. V. Zarkov, "Combining the modal expansion and dipole equivalence methods for coated plasmonic particles of various shapes," J. Phys. Chem. C 129, 10958-10974 (2025).

[4] R. Yu, L. M. Liz-Marzán, and F. J. García de Abajo, "Universal analytical modeling of plasmonic nanoparticles," Chem. Soc. Rev. 46, 6710–6724 (2017).

[5] N. G. Khlebtsov, "T-matrix method in plasmonics: an overview," J. Quant. Spectrosc. Radiat.Transfer 123, 184–217 (2013).

[6] J.-F. Masson, J. S. Biggins, and E. Ringe, "Machine learning for nanoplasmonics," Nat. Nanotechnol. 18, 111–123 (2023).

[7] A. J. Canning, J. Q. Li, J. Chen, K. Hoang, T. Thorsen, A. Vaziri, and T. Vo-Dinh, "Tunable and scalable production of nanostar particle platforms for diverse applications using an AI-integrated automated synthesis system," J. Mater. Sci. 60, 3768-3785 (2025).

[8] D. Schletz, M. Breidung, and A. Fery, "Validating and utilizing machine learning methods to investigate the impacts of synthesis parameters in gold nanoparticle synthesis," J. Phys. Chem. C 127, 2, 1117–1125 (2023)

[9] U. Kreibig and M. Vollmer, *Optical Properties of Metal Clusters* (Springer-Verlag, Berlin, 1995).

[10] P. Drude, "Über die Bezihung der Dielectricitätconstanten zem optischen Brechungsexponenten," Ann. der Physik 284 (3), 536-545 (1893).

[11] Liz-Marzán L. (Ed.). *Colloidal Synthesis of Plasmonic Nanometals* (Jenny Stanford Publishing, New York, 2021).

[12] J. L. Montaño-Priede, A. Sanchez-Iglesias, S. A. Mezzasalma, J. Sancho-Parramon, and M. Grzelczak, "Quantifying shape transition in anisotropic plasmonic nanoparticles through geometric inversion. application to gold bipyramids," J. Phys. Chem. Lett. 15, 3914−3922 (2024).

[13] R. Fuchs, "Theory of the optical properties of ionic crystal cubes," Phys. Rev. B 11, 1732 – 1740 (1975).

[14] T. J. Davis and D. E. Gómez, "Colloquium: an algebraic model of localized surface plasmons and their interactions," Rev. Mod. Phys. 89, 011003 (2017).

[15] M. Meier and A. Wokaun. "Enhanced fields on large metal particles: dynamic depolarization," Opt. Lett. 8, 581–583 (1983).

[16] A. Moroz, "Depolarization field of spheroidal particles," J. Opt. Soc. Am. B 26, 517–527 (2009).

[17] J. D. Jackson, *Classical Electrodynamics*, 3rd ed. (Wiley, New York, 1999).

[18] N. G. Khlebtsov, "Extinction and scattering of light by nonshperical particles in absorbing media," J. Quant. Spectr. Radiat. Transfer 280, 108069 (2022).

[19] C. F. Bohren and D. R. Huffman, *Absorption and Scattering of Light by Small Particles* (Wiley, New York, 1983).

[20] W. T. Doyle, "Absorption of light by colloids in alkali halide crystals," Phys. Rev. 111, 1067 (1958).

[21] W. Hampe, "Beitrag zur Deutung der anomalen optischen Eigenschaften feinstteiliger Metallkolloide in groβer Konzentration. Teil II: Experimentelle Ermittlung der Absorptionskurve und Deutung des Absorptionsmechanismus des Systemes Gold-Si0$_2$", Z. Physik, 152, 476–494 (1958).

[22] H. Römer and C. v. Fragstein, "Bestimmung des Absorptionskoeffizienten und des Brechungsquotienten von kolloidalem Gold. Ein Beitrag zur "Anomalie der optischen Konstanten," Z. Physik 163, 27–43 (1961).

[23] C. v. Fragstein and H. Romer, "Uber die Anomalie der optischen Konstanten," Z. Physik 151, 54–71 (1958).

[24] U. Kreibig and C. v. Fragstein, "The limitation of electron mean free path in small silver particles," Z. Phys. 324, 307–323 (1969).

[25] R. H. Doremus, "Optical properties of small gold particles," J. Chem. Phys. 40, 2389–2396 (1964).

[26] U. Kreibig, "Electronic properties of small silver particles: the optical constants and their temperature dependence," J. Phys. F: Met. Phys. 4, 999–1014 (1974).

[27] C. G. Granqvist and O. Hunderi, "Optical properties of ultrafine gold particles," Phys. Rev. B, 16, 3513–3534 (1977).

[28] N. G. Khlebtsov, V. A. Bogatyrev, L. A. Dykman, and A. G Melnikov, "Spectral extinction of colloidal gold and its biospecific conjugates," J. Colloid Interface Sci., 180, 436–445 (1996).

[29] U. Kreibig and L. Genzel, "Optical absorption of small metal particles," Surf. Sci. 156, 678–700 (1985).

[30] N. G. Khlebtsov, "Optics and biophotonics of nanoparticles with a plasmon resonance," Quant. Electron. 26, 504-529 (2008).

[31] A. Kawabata and R. Kubo, "Electronic properties of fine metallic particles. II. Plasma resonance absorption," J. Phys. Soc. Japan, 21, 1765–1772 (1966).

[32] A. A. Lushnikov and A. J. Simonov, "Surface plasmons in small metal particles," Z. Physik 270, 17 (1974)

[33] R. Ruppin and H. Yatom, "Size and shape effects on the broadening of the plasma resonance absorption in metals," Phys. Stat. Sol. (B) 74, 647-654 (1976)

[34] L. Genzel, T. P. Martin, and U. Kreibig, "Dielectric function and plasma resonances of small metal particles," Z. Physik B 21, 339–346 (1975).

[35] H. Hövel, S. Fritz, A. Hilger, U. Kreibig, and M. Vollmer, "Width of cluster plasmon resonances: Bulk dielectric functions and chemical interface damping," Phys. Rev. B 48, 18178–18188 (1993).

[36] B. Foerster, A. Joplin, K. Kaefer, S. Celiksoy, S. Link, and C. Sönnichsen, "Chemical interface damping depends on electrons reaching the surface," ACS Nano 11, 2886–2893 (2017).

[37] A. Wokaun, J. P. Gordon, and P. F. Liao, "Radiation damping in surface-enhanced Raman scattering," Phys. Rev. Lett. 48, 957-960 (1982).

[38] C. Sönnichsen, T. Franzl, T. Wilk, G. von Plessen, J. Feldmann, O. Wilson, and P. Mulvaney, "Drastic reduction of plasmon damping in gold nanorods," Phys. Rev. Lett. 88, 077402 (2002).

[39] E. A. Coronado and G. C. Schatz, "Surface plasmon broadening for arbitrary shape nanoparticles: A geometrical probability approach," J. Chem. Phys. 119, 3926-3934 (2003).

[40] N. G. Khlebtsov, S. V. Zarkov, V.A. Khanadeev, and Y. A. Avetisyan, "Novel concept of two-component dielectric function for gold nanostars: theoretical modelling and experimental verification," Nanoscale 12, 19963–19981 (2020).

[41] B. Khlebtsov, V. Khanadeev, T. Pylaev, and N. Khlebtsov, "A new T-Matrix solvable model for nanorods: TEM-based ensemble simulations supported by experiments," J. Phys. Chem. C 115, 6317–6323 (2011).

[42] S. Berciaud, L. Cognet, P. Tamarat, and B. Lounis, "Observation of intrinsic size effects in the optical response of individual gold nanoparticles" Nano Lett. 5, 515–518 (2005).

[43] C. Novo, D. Gomez, J. Perez-Juste, Z. Zhang, H. Petrova, M. Reismann, P. Mulvaney, and G. V. Hartland, "Contributions from radiation damping and surface scattering to the linewidth of the longitudinal plasmon band of gold nanorods: a single particle study," Phys. Chem. Chem. Phys. 8, 3540–3546 (2006).

[44] M. Hu, C. Novo, A. Funston, H. Wang, H. Stalev, S. Zou, P. Mulvaney, Y. Xia, and G. V. Hartland, "Dark-field microscopy studies of single metal nanoparticles: understanding the factors that influence the linewidth of the localized surface plasmon resonance," J. Mater. Chem. 18, 1949–1960 (2008).

[45] A. D. Kondorskiy and A. V. Mekshun, "Effect of geometric parameters of metallic nanoprisms on the plasmonic resonance wavelength," J. Russ. Laser Res. 44 (6), 627–636 (2023).

[46] R. L. Olmon, B. Slovick, T. W. Johnson, D. Shelton, S.-H. Oh, G. D. Borema, and M. B. Raschke, "Optical dielectric function of gold," Phys. Rev. B 86, 235147 (2012).

[47] H. Chen, L. Shao, Q. Li, and J. Wang, "Gold nanorods and their plasmonic properties," Chem. Soc. Rsev. 42, 2679–2724 (2013).

[48] V. V. Tuchin, *Tissue Optics: Light Scattering Methods and Instruments for Medical Diagnosis,* 3rd ed. (SPIE, 2015), DOI: 10.1117/3.1003040.

[49] H. Chen, X. Kou, Z. Yang, W. Ni, and J. Wang, "Shape- and size-dependent refractive index sensitivity of gold nanoparticles," Langmuir 24, 5233-5237 (2008).

[50] T. H. Chow, N. Li, X. Bai, X. Zhuo, L. Shao, and J. Wang, "Gold nanobipyramids: an emerging and versatile type of plasmonic nanoparticles," Acc. Chem. Res. 52, 2136−2146 (2019).

[51] J.-H. Lee, K. J. Gibson, G. Chen, and Y. Weizmann, "Bipyramid-templated synthesis of monodisperse anisotropic gold nanocrystals," Nat. Commun. 6, 7571 (2015).

[52] D. Chateau, A. Desert, F. Lerouge, G. Landaburu, S. Santucci, and S. Parola, "Beyond the concentration limitation in the synthesis of nanobipyramids and other pentatwinned gold nanostructures," ACS Appl. Mater. Interfaces 11, 39068−39076 (2019).

[53] A. L. Nguen, Q. J. Griffin, A. Wang, S. Zou, and H. Jing, "Optimization of the surfactant ratio in the formation of penta-twinned seeds for precision synthesis of gold nanobipyramids with tunable plasmon resonances," J. Phys. Chem. C 129, 4303−4312 (2025).

[54] A. Sánchez-Iglesias and M. Grzelczak, "Expanding chemical space in the synthesis of gold bipyramids," Small, 21, 2407735 (2025).

[55] M. Liu and P. Guyot-Sionnest, "Mechanism of silver(I)-assisted growth of gold nanorods and bipyramids," J. Phys. Chem. B 109, 22192−22200 (2005).

[56] N. G. Khlebtsov, "Optical models for conjugates of gold and silver nanoparticles with biomacromolecules", J. Quant. Spectrosc. Radiat. Transfer 89, 143–153 (2004).

[57] H. Jing, N. Large, Q. Zhang, and H. Wang, "Epitaxial growth of $Cu_2O$ on Ag allows for fine control over particle geometries and optical properties of Ag–$Cu_2O$ core–shell nanoparticles," J Phys. Chem. C 118, 19948–19963 (2014).

# Supplementary Material

# Analytical models for coated plasmonic particles: effects of shape and size-corrected dielectric function

Nikolai G. Khlebtsov[1,2,*], Sergey V. Zarkov[1,3]

[1]Institute of Biochemistry and Physiology of Plants and Microorganisms, "Saratov Scientific Centre of the Russian Academy of Sciences," 13 Prospekt Entuziastov, Saratov 410049, Russia

[2]Saratov State University, 83 Ulitsa Astrakhanskaya, Saratov 410012, Russia

[3]Institute of Precision Mechanics and Control, "Saratov Scientific Centre of the Russian Academy of Sciences," 24 Ulitsa Rabochaya, Saratov 410028, Russia

[*]To whom correspondence should be addressed. E-mail: (NGK) khlebtsov@ibppm.ru

**Section S1. MEM parameters for selected particle shapes**

For rods and discs, MEM parameters coincide with those given in Ref.[1] For coated rods and disks their geometrical sizes should be increased by twice the shell thickness $2s$. Accordingly, the aspect ratios of coated rods and disks are $AR_2 = \dfrac{L_1 + 2s}{d_1 + 2s}$ and $AR_2 = \dfrac{D_1 + 2s}{t_1 + 2s}$, respectively. Below, the symbols $h_1$ and $h_2$ correspond to the generalized aspect ratios $AR_1$ for the metal core and $AR_2$ for the shell. It should be emphasized here that we use the same MEM functions for metal cores and shells due to DEM philosophy. Below, we provide analytical expressions for MEM parameters $\eta_1$, $a_{12}$, $a_{14}$, $V_1^m$, and for the initial and rounded particle volumes $V$ and $V_r$ without derivation. For rods and disks $V = V_r$. All details can be found in Ref.[2]

**S1.1 Rods**

$$\eta_1(h) = -1.73 h^{1.45} - 0.296,\ a_{12}(h) = 6.92/\left[1 - \eta_1(h)\right],\ a_{14}(h) = -11 h^{-2.49} - 0.0868, \quad \text{(S1)}$$

$$V_1^m / V_{NR} = 0.896, \quad V_{NR} / L_1^3 = \frac{\pi(3h-1)}{12h^3}. \tag{S2}$$

**S1.2 Disks**

$$\eta_1(h) = -1.36 h^{0.872} - 0.479, \quad a_{12}(h) = 7.05 / [1 - \eta_1(h)], \quad a_{14}(h) = -10.9 h^{-0.98}, \tag{S3}$$

$$V_1^m / V_{ND} = 0.944, \quad V_{ND} / L_1^3 = \frac{\pi}{24 h^3}[4 + 3(h-1)(2h + \pi - 2)]. \tag{S4}$$

**S1.3 Triangle prism**

The generalized aspect ratios of bare and coated nanoprisms are characterized by similar expressions in $AR_i \equiv h_i = L_i / t_i$, $i = 1, 2$, where the indices 1 and 2 stand for the metal core and shell, respectively; $L_2 = L_1 + 2\sqrt{3} s$, and $t_2 = t_1 + 2s$. Here, $t_1$ and $t_2$ designate the bare and coated prism thicknesses, respectively. Note that both lengths here are given in terms of unrounded particles. For rounded particles with a rounding radius[2] $r = (1 + 1/AR) \times L / 40$, the MEM parameters are:

$$\eta_1(h) = -3.2483 - 1.2047 h + 0.00632112 h^2, \tag{S5}$$

$$a_{12}(h) = 5.57 / [1 - \eta_1(h)], \quad a_{14}(h) = -6.83 / [1 - \eta_1(h)], \tag{S6}$$

$$V_1^m / V_{NT} = 0.55 + 0.155 \left[1 - e^{-(h-2)/5}\right], \tag{S7}$$

$$V_{NTr} / L_1^3 = \frac{\sqrt{3}}{4h} - \frac{(h+1)^2}{1600 h^2} \left( \frac{3\sqrt{3} - \pi}{h} + 6(1 - \tfrac{1}{4}\pi) \right). \tag{S8}$$

In all the above equations, the symbol $h$ stands for the aspect ratio of the rounded core ($AR_1$) or shell ($AR_2$), respectively. The volume of a rounded coated nanoprism can be calculated by the above Eq. (S8) with $L \equiv L_2 = L_1 + 2\sqrt{3} s$ and $h = AR_2 = L_2 / (t_1 + 2s)$.

**S1.4 Rounded bicone**

**S1.4.1 Factorized MEM model for bicones with small tip curvature radii.**

A detailed description of all geometrical parameters is given in Ref.[2] Specifically, for numerical simulations of bicones with sharp tips, we used small rounding radii $r_i = L_i/40 AR_i = d_i/40, i=1,2$ for the core and shell tips and $R_i = d_i/100, i=1,2$ for the core and shell bases. For metal bicones, this parametrization was used by Yu et al.[1] The size and shape of the initial bicone (unrounded metal core) are characterized by the total $L_1$ length, diameter $d_1$, and aspect ratio $AR_1 = L_1/d_1$. The unrounded shell bicone is characterized by similar parameters $L_2$, $d_2$, and $AR_2 = L_2/d_2 = AR_1$ with

$$L_2 = L_1 + 2s/\sin\alpha = L_1 + 2s\sqrt{1+AR_1^2}, \tag{S9}$$

$$d_2 = d_1 + 2s/\cos\alpha = d_1 + 2s\sqrt{1+AR_1^2}/AR_1. \tag{S10}$$

$$\alpha = \arctan\frac{1}{AR_1}. \tag{S11}$$

The rounding radii of the core tips and base are $r_1$ and $R_1$, respectively. The rounded length and diameter of the core are $L_{1r}$ and $d_{1r}$, respectively. For the coated cone, we have $L_{2r} = L_{1r} + 2s$ and $d_{2r} = d_{1r} + 2s$. Explicit relations for these parameters are given in Ref.[2] The volume of the initial bicone with the total length $L_i$ and diameter $d_i$ equals

$$V_{BCi} = \frac{2}{3}\pi(d_i/2)^2\frac{L_i}{2} = \frac{\pi L_i^3}{12AR_i^2}, \quad AR_i = L_i/d_i, i=1,2, \tag{S12}$$

where the indices 1 and 2 stand for the core and shell, respectively. For a rounded bicone, we have to subtract from the volume (S12) the volume of the rounded tips[2]

$$V_{iexc}^{tip} = \frac{2\pi}{3}r_i^3\frac{(1-\sin\alpha)^2}{\sin\alpha}, \quad i=1,2 \tag{S13}$$

where $r_2 = r_1 + s$, and the volume of the rounded base,[2]

$$V_{iexc}^{base} = 2\pi R^2\left\{\frac{d_i}{2}(\tan\alpha - \alpha) + R_i\left[-\frac{\sin\alpha}{3}\left(1+\frac{2}{\cos^2\alpha}\right)+\frac{\alpha}{\cos\alpha}\right]\right\}. \tag{S14}$$

S3

Note that the angle $\alpha$ is the same for the core and shell. In Eq. (S14), the rounding radius $R_i$ can be $R_1$ for the core and $R_2 = R_1 + s$ for the shell. Finally, the rounded bare and coated bicones volume is calculated as the difference between the initial and excluded base and tip volumes

$$V_{iBCr} = V_{iBC} - \left(V_{iexc}^{tip} + V_{iexc}^{base}\right). \tag{S15}$$

The above analytical expressions for the rounded bicone volume were checked with a direct numerical integration by COMSOL.

To calculate the modal volumes (mode 1) for Au bicones, we used the following factorization model

$$V_1^m = V_{BCr} \times f_{Vh}(h_r) f_{Vs}(s), \tag{S16}$$

$$f_{Vh}(x) = 0.32259 + 0.086858x - 0.015968x^2 + 0.0013815x^3 - 4.55693\text{E-}05x^4, \tag{S17}$$

$$f_{Vs}(x) = \begin{cases} 1, & s \leq 0.4 \\ 0.92 - 0.0398176x + 0.00660429x^2 + 0.00543648x^3 - 0.000985436x^4, & x = \ln(s) \end{cases} \tag{S18}$$

The MEM parameters $\eta_j^p(h_r)$ for the first mode $j = 1$ and longitudinal excitation along the bicone symmetry axis are calculated by the following factorized approximation

$$\eta_1 = -f_{\eta h}(h_r) \times f_{\eta s}(s), \tag{S19}$$

$$f_{\eta h}(x = h_r) = 0.29105 + 2.6267x + 0.84365x^2 - 0.0057085x^3 - 0.00043102x^4, \tag{S20}$$

$$f_{\eta s}(s) = \begin{cases} 1, & s = 0, \\ 1.29766 s^{0.121111}, & 1 \leq s \leq 30 \end{cases}. \tag{S21}$$

Finally, the last two MEM parameters for the first longitudinal mode are[1]

$$\alpha_{12}(h_r) = \frac{1.34}{1-\eta_1}, \quad \alpha_{14}(h_r) = -\frac{1.04}{1-\eta_1}. \tag{S22}$$

In all the above equations, $h_r$ stands for the aspect ratio of the rounded core ($AR_{1r}$) or shell ($AR_{2r}$), respectively.

For Ag bicones, the above MEM parameters are as follows:

$$f_{Vh}(x) = 0.655276 - 0.221254x + 0.0971170x^2 - 0.0174292x^3 + 0.00112386x^4, \qquad (S23)$$

$$f_{Vs}(x) = \begin{cases} 1, & s = 0, \\ \exp\left[-\left(0.0833816 + 0.0541018x - 0.0230390x^2 - 0.00114804x^3\right)\right], \\ x = \ln(s), \end{cases} \qquad (S24)$$

$$f_{\eta h}(x = h_r) = 0.8330985 + 2.217802x + 1.00832x^2 - 0.0332392x^3 - 0.00120781x^4, \qquad (S25)$$

$$f_{\eta s}(s) = \begin{cases} 1, & s = 0, \\ 1. + 0.229102 + 0.203835x - 0.0178395x^2 \\ -0.0126139x^3 + 0.00135825x^4, & x = \ln s. \end{cases} \qquad (S26)$$

If the rounding radii of core and base are large enough, the functions $f_{Vs}(s)$ and $f_{\eta s}(s)$ can be replaced with 1, and the functions $f_{Vh}(x)$ and $f_{\eta h}(x)$ should be properly defined from the fitting to numerical data. In other words, the factorized MEM representations (S16) and (S19) are reduced to standard MEM ones (see below Section 1.4.2).

**S1.4.2 MEM model for bicones with nanometer-sized tip curvature.**

It follows from experimental data[3] that the dependence of the tip curvature radius $R_{tip}$ on the rounded length $L_r$ can be approximated by the relation $R_{tip} = -0.15L_r + 23$, for experimental series 1-2 and by $R_{tip} = -0.17L_r + 13$ for experimental series 3-4. Considering the average diameter close to 30 nm for particles of series 1-2, we finally get

$$R_{tip} = -5AR_r + 23, \quad AR_r \leq 4.5. \qquad (S27)$$

In agreement with the data of Ref.[3] we assume in this model that the base rounding radius is zero.

We considered rounded bicones with aspect ratios $2 \leq AR_r \leq 4$; therefore, the rounding radii varied within the range $3 \leq R_{tip}(\text{nm}) \leq 13$. For such nanometer-sized radii, the factorized MEM version is reduced to the usual MEM model and Eqs. (S16) and (S19) are replaced with

$$V_1^m / V_{BCr} = f_{Vh}(h_r), \qquad (S28)$$

$$\eta_1 = -f_{\eta h}(h_r).  \quad (S29)$$

Functions in the right parts of Eqs. (S28) and (S29) were calculated using cubic splines. The spline points for gold and silver bicones are

|  | Au | | Ag | |
|---|---|---|---|---|
| $h_r$ | $f_{\eta h}(h_r)$ | $f_{Vh}(h_r)$ | $f_{\eta h}(h_r)$ | $f_{Vh}(h_r)$ |
| 2 | 5.735 | 0.963 | 5.950 | 0.97 |
| 3 | 11.16 | 0.970 | 11.41 | 0.97 |
| 4 | 19.69 | 0.795 | 19.85 | 0.80 |
| 4.5 | 29.20 | 0.450 | 29.20 | 0.46 |

**S1.5 Rounded bipyramid**

The geometrical parameters of our rounded bipyramid (BP) model are based on the embedded bicone model. This model is characterized by the diameter of a circle inscribed in the pentagon of the BP base and its height, which coincides with the height of the BP. It follows from experimental data (e.g., Ref.[3] and references therein) that the tip rounding radius $R_{tip}$ increases with a decrease in the actual BP length $L_{1r}$. Similarly, the base rounding radius $R_{base}$ also increases during the chemical etching. Unfortunately, in the theoretical model, the length of a rounded BP and its aspect ratio depend on the rounding radii, so we have a nonlinear relationship. To avoid this difficulty, we assume the linear relations between the rounding radii and the aspect ratio of the initial (unrounded) BP

$$R_{tip}(\text{nm}) = 2 + 2(6 - AR_1), 1 \leq AR_1 \leq 6, \quad (S30)$$

$$R_{base}(\text{nm}) = 16 - 2AR_1. \quad (S31)$$

In our simulations, we fixed the inscribed bicone diameter as the BP width and varied the aspect ratio from 2 to 6. This resulted in a progressive decrease in the rounding radii, consistent with the experimental measurements presented in Ref.[3]

Given the bipyramid diameter, aspect ratio, and rounding radii (S30) and (S31), we numerically calculated the rounded BP volume and all geometrical parameters of the embedded bicone analytically. Then, we approximate the BP volume by a polynomial of the rounded aspect ratio

$$V_{BPr}(s=0) = -7486.03 + 18338.49 AR_{1r} - 1433.54 AR_{1r}^2, \quad (S32)$$

where the shell thickness $s$ is assumed to be zero. Then, for a given shell thickness $s$, we calculated analytically the ratio $F_V(s) = V_{BCr}(s)/V_{BCr}(s=0)$. Finally, the volume of a rounded coated BP was calculated as $V_{BCr}(s) = F_V(s)V_{BCr}(s)/V_{BCr}(s=0)$. The accuracy of this approximation is better than 0.002%.[2]

With the rounded BP volume in hand, we calculated other MEM parameters as follows:

$$\begin{aligned} \eta_1(h_r) &= -f_\eta(h_r), \\ f_\eta(x) &= -1.36762 + 3.0204378x + 0.09294389x^2 + 0.07895368x^3 \end{aligned} \quad (S33)$$

$$a_{12}(h_r) = 5.57/[1-\eta_1(h_r)], \quad a_{14}(h_{ir}) = -6.83/[1-\eta_1(h_{ir})], \quad (S34)$$

$$\begin{aligned} V_1^m &= f_V(h_r)V_{BPr}, \\ f_V(x) &= 0.976704 - 0.0237377x - 0.00453238x^2 \end{aligned} \quad (S35)$$

For silver bipyramids, Eqs. (S33)-(S35) should be modified as follows:

$$\begin{aligned} \eta_1(h_r) &= -f_\eta(h_r), \\ f_\eta(x) &= -0.323882 + 1.42211x + 1.01670x^2 - 0.129240x^3 + 0.0160452x^4 \end{aligned} \quad (S36)$$

$$\begin{aligned} V_1^m &= f_V(h_r)V_{BPr}, \\ f_V(x) &= 0.519999 + 0.562648x - 0.279774x^2 + 0.0561240x^3 - 0.00413525x^4 \end{aligned} \quad (S37)$$

In all the above equations, the symbol $h_r$ stands for the aspect ratio of the rounded core ($AR_{1r}$) or shell ($AR_{2r}$), respectively.

**Section S2. Effective scattering length of electrons**

The effective scattering length in the billiard scattering model by Coronado and Schatz[4] is proportional to the ratio of the particle volume to its surface area $l_s = 4V/S$. Below, we provide explicit expressions for several particle models used in this work,

**2.1 Rod with semispherical ends**

$$l_s = d\left(1 - \frac{1}{3h}\right), \tag{S38}$$

where $h \equiv AR = \frac{L}{d}$ is the aspect ratio. For spheres, $h = 1$ and $l_s = 2d/3$. For a long rod $l_s = d$.

**2.2 Disk with circular edges**

The disk diameter, thickness, and aspect ratio are $D$, $t$, and $h = D/t$, respectively (Figure S1). The disk volume is

$$V = \frac{\pi t^3}{24}\left[4 + 3(h-1)(\pi + 2h - 2)\right]. \tag{S39}$$

Below we provide a short derivation.

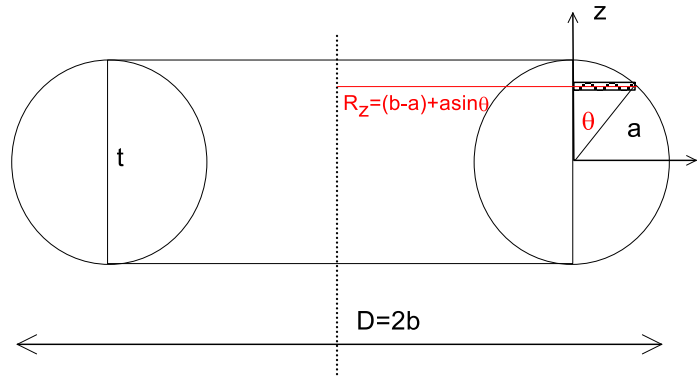

**Figure S1**. Schematics for derivation of a disk volume formula.

$$V = \pi \frac{1}{4}(D-t)^2 t + V_1 = \pi t^3 (h-1)^2 + V_1$$

$$V_1 = 2\int_0^a \pi\left(R_z^2 - (b-a)^2\right)dz = 2\pi\int_0^a \left(a^2 - z^2 + 2(b-a)\sqrt{a^2 - z^2}\right)dz = \frac{4}{3}\pi a^3 + (b-a)a^2\pi^2$$

$$V_1 = \frac{4}{3}\pi a^3 + (b-a)a^2\pi^2 = \frac{1}{6}\pi t^3 + \frac{1}{8}(h-1)t^3\pi^2$$

S8

$$V = \pi \frac{1}{4}(D-t)^2 t + V_1 = \frac{\pi}{4}t^3(h-1)^2 + \frac{\pi}{6}t^3 + \frac{\pi^2}{8}(h-1)t^3 =$$

$$= \frac{\pi t^3}{6}\left[1 + \frac{3}{2}(h-1)^2 + \frac{3}{4}\pi(h-1)\right] = \frac{\pi t^3}{6}\left[1 + \frac{3}{4}(h-1)(\pi + 2h - 2)\right]$$

The disk surface is:

$$S = 2\int_0^{\pi/2} r d\theta (r\sin\theta + R)2\pi = 4\pi r^2(1 + \pi R/2r) + 2\frac{\pi(D-t)^2}{4} = \pi t^2\left[1 + \frac{h-1}{2}(\pi + h - 1)\right],$$

where $R = (D-t)/2$ and $r = t/2$. Finally, we have

$$l_s = t \times f_D(h), \qquad (S40)$$

$$f_D(h) = \frac{2}{3}\frac{1 + \frac{3}{4}(h-1)(\pi + 2h - 2)}{1 + \frac{h-1}{2}(\pi + h - 1)}, \qquad (S41)$$

For spheres $f_D = 1$ and again $l_c = 2d/3$. For a thin disk $l_c = 2t$. A disk of such kind cannot be thicker than a sphere with $D = t$.

### 2.3 Triangle prism

Because of the small rounding radii of tips and base, we can calculate the scattering length for the initial unrounded prism:

$$V = S_{base} t = \frac{L^2\sqrt{3}}{4h}L, \qquad (S42)$$

$$S = 2\frac{L^2\sqrt{3}}{4} + 3Lt = \frac{L^2\sqrt{3}}{2} + \frac{3L^2}{h} = \frac{L^2\sqrt{3}}{2}(1 + \frac{2\sqrt{3}}{h}), \qquad (S43)$$

$$l_s = \frac{4V}{S} = \frac{4\frac{L^2\sqrt{3}}{4h}L}{\frac{L^2\sqrt{3}}{2}(1 + \frac{2\sqrt{3}}{h})} = \frac{2t}{(1 + \frac{2\sqrt{3}}{h})}, \qquad (S44)$$

where $h \equiv AR = L/t$. For a thin prism $l_s = 2t$, similarly to a thin disk formula. It is a specific feature of the billiard model.[4] For a thick prism (which is close to a rod), we have $h = AR = L/t \to 0$, so that

S9

$$l_s = \frac{2t}{1+\frac{2\sqrt{3}}{h}} \approx \frac{th}{\sqrt{3}} = \frac{L}{\sqrt{3}}. \qquad (S45)$$

The effective length is proportional to the side length divided by the square root of 3. Note that for a long rod, the effective length is proportional to its diameter.

## 2.4 Bicone

As the rounding radii of tips and base are small, we can calculate the scattering length for an unrounded bicone. The bicone volume is

$$V = \frac{2}{3}\pi(d/2)^2 \frac{L}{2} = \frac{\pi L^3}{12h^2} = \frac{\pi d^3 h}{12}, \qquad (S46)$$

where $h = L/d$. The bicone surface equals

$$S = 2\times\pi(d/2)\sqrt{(d/2)^2 + (L/2)^2} = \frac{\pi d^2}{2}\sqrt{1+h^2}, \qquad (S47)$$

whence

$$l_s = 4V/S = \frac{\frac{\pi d^3 h}{3}}{\frac{\pi d^2}{2}\sqrt{1+h^2}} = \frac{2}{3}d\frac{h}{\sqrt{1+h^2}}. \qquad (S48)$$

Sometimes, the equivolume sphere radius

$$R_{ev} = \sqrt[3]{\frac{3}{4\pi}\frac{\pi d^3 h}{12}} = \frac{d}{2}\sqrt[3]{\frac{h}{2}}, \qquad (S49)$$

is used as an effective scattering length. As for spheres $l_s = \frac{4}{3}R$, one has to use a multiplier ¾ for $l_s$ of other shapes when comparing the size-limiting effects with spheres of a radius R with the same scattering constant $A_s$. Alternatively, one has to use different scattering constants $A_s$ to make the physical dumping equivalent.

## 2.4 Bipyramid

According to Ref.[3], the optical properties of a pentagonal bipyramid can be adequately modeled by an inscribed bicone with the same total length $L_b = L$ and an inscribed diameter $d_b$. Then, we can use the above Eq. (48) with $h_b = L_b / d_b$.

**Section S3. Volume fraction of the local field**

**3.1 Nanorods**

The local field near NR tips is assumed to be localized between two tip semispheres with diameters of $d$ and $d + 2l_{NR}$ (Figure S2 and Figure 9C of the main text). Then, the localized field volume is

$$V_l = \frac{\pi}{6}\left[(d + 2l_{NR})^3 - d^3\right] \approx \pi d^2 l_{NR}. \tag{S50}$$

Thus, the localized field volume fraction is

$$\varphi_l^{NR} = \frac{V_l}{V_1} = \frac{4l_{NR}}{L(1 - 1/3h_1)} \approx \frac{4l_{NR}}{L}, \tag{S51}$$

where a typical condition $h_1 \geq 2$ has been considered.

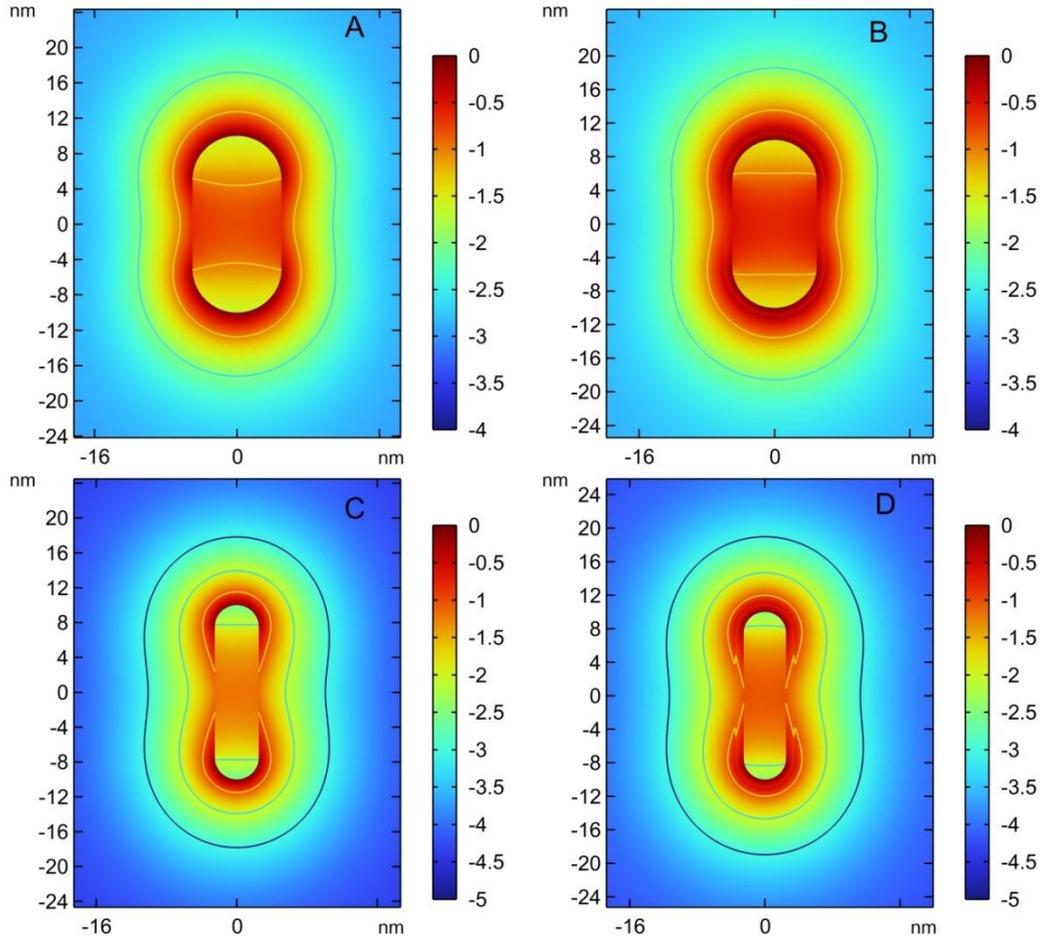

**Figure S2.** The local field distribution $\ln(E/E_{max})$ around nanorod tips, where $E_{max}$ is the field module at the tip surface at the resonance wavelength. Calculations for bare (A, C) and coated (B, D) gold nanorods with a length of 20 nm and aspect ratios of 2 (A, B) and 4 (C, D). The dielectric shell thickness is 1 nm. Shown are lines where the field amplitude is decreased by $e^{-n}$ for $n=1$, 2, and 3. In contrast to the gold bicone with the same length and aspect ratios (Figure 9 of the main text), the localization lengths for isolines $n=1$ are one order greater, ranging from 2 nm (panels C, D) to 4 nm (panel A, B).

### 3.2 Bicones

In the first approximation, we calculate the localized field volume $V_l$ as the difference between the rounding tip semisphere of a radius $R_{tip}$ and that of an increased radius $R_{tip}+l_{BC}$ with $l_{BC} \sim 0.2$ nm (Figure 9B of the main text). Taking into account both tip volumes, we obtain

$$V_l = \frac{4\pi}{3}\left[\left(R_{tip}+l_{BC}\right)^3 - R_{tip}^3\right] \approx 4\pi R_{tip}^2 l_{BC}, \tag{S52}$$

where the condition $R_{tip} \gg l_{BC}$ has been taken into account. By using an approximate expression for the volume of a rounded bicone [2]

$$V_1 \approx \frac{\pi d_1^2 L}{12}, \tag{S53}$$

and the rounding radius value $R_{tip} = d_1 / 40$, we arrive at the final result

$$\varphi_l^{BC} = \frac{V_l}{V_1} = 48 \left(\frac{R_{tip}}{d_1}\right)^2 \frac{l_{BC}}{L} = 3 \times 10^{-2} \frac{l_{BC}}{L}. \tag{S54}$$

Considering Eqs. (S51) and (S54), we finally get

$$\varphi_l^{BC} / \varphi_l^{NR} = 0.75 \times 10^{-2} \frac{l_{BC}}{l_{NR}}. \tag{S55}$$

**Section S4. Additional Figures**

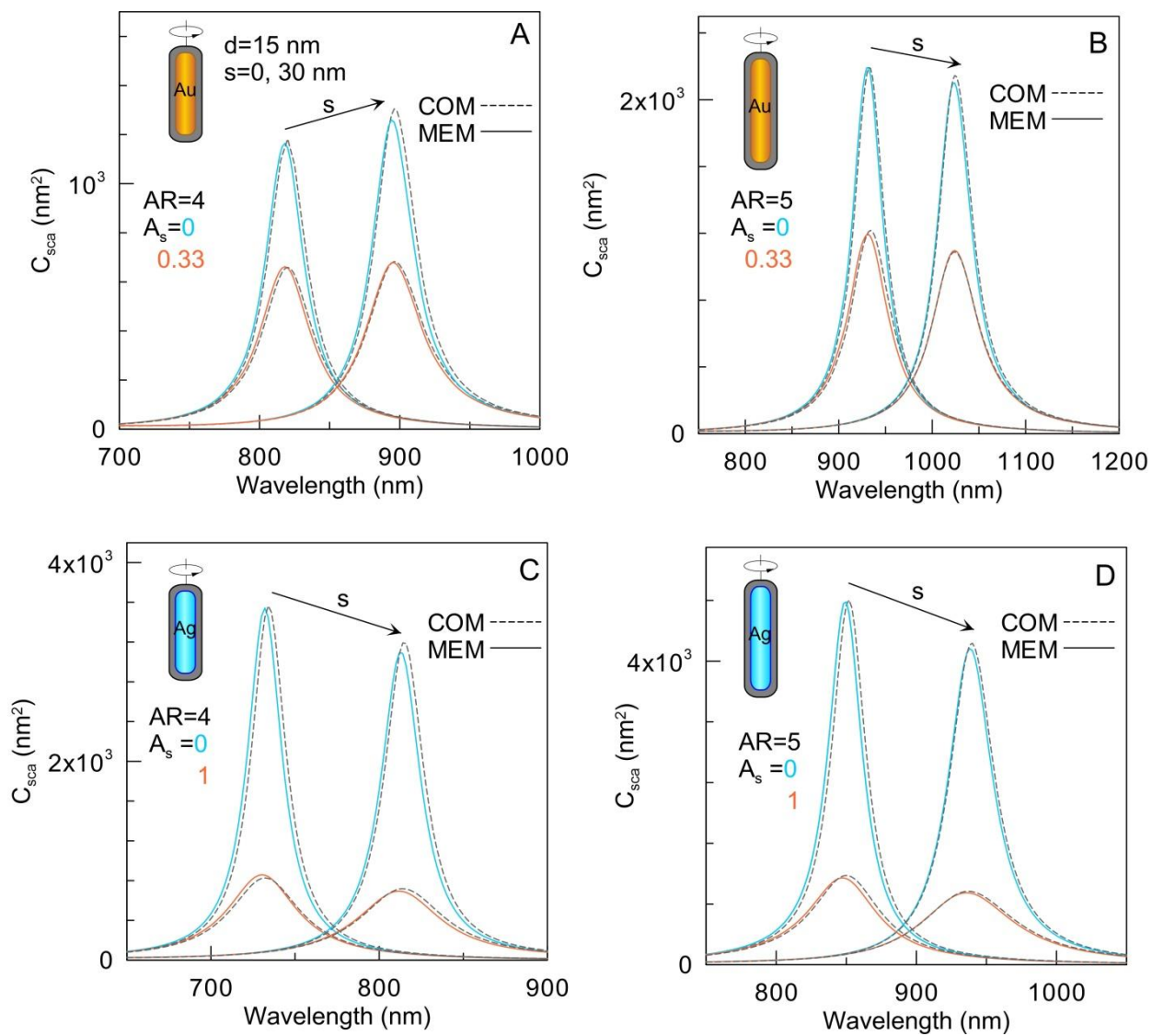

**Figure S3**. Scattering spectra of randomly oriented gold (A, B) and silver (C, D) bare ($s=0$) and coated ($s=30$)) nanorods with aspect ratios of 4 (A, C) and 5 (B, D) and a fixed diameter of 15 nm. Calculations by COMSOL (dashed lines) and MEM (solid lines) for bulk ($A_s = 0$, blue) and size-corrected ($A_s = 0.33$ for gold and $A_s = 1$ for silver, orange) optical constants. The electron scattering constants are indicated in the panels.

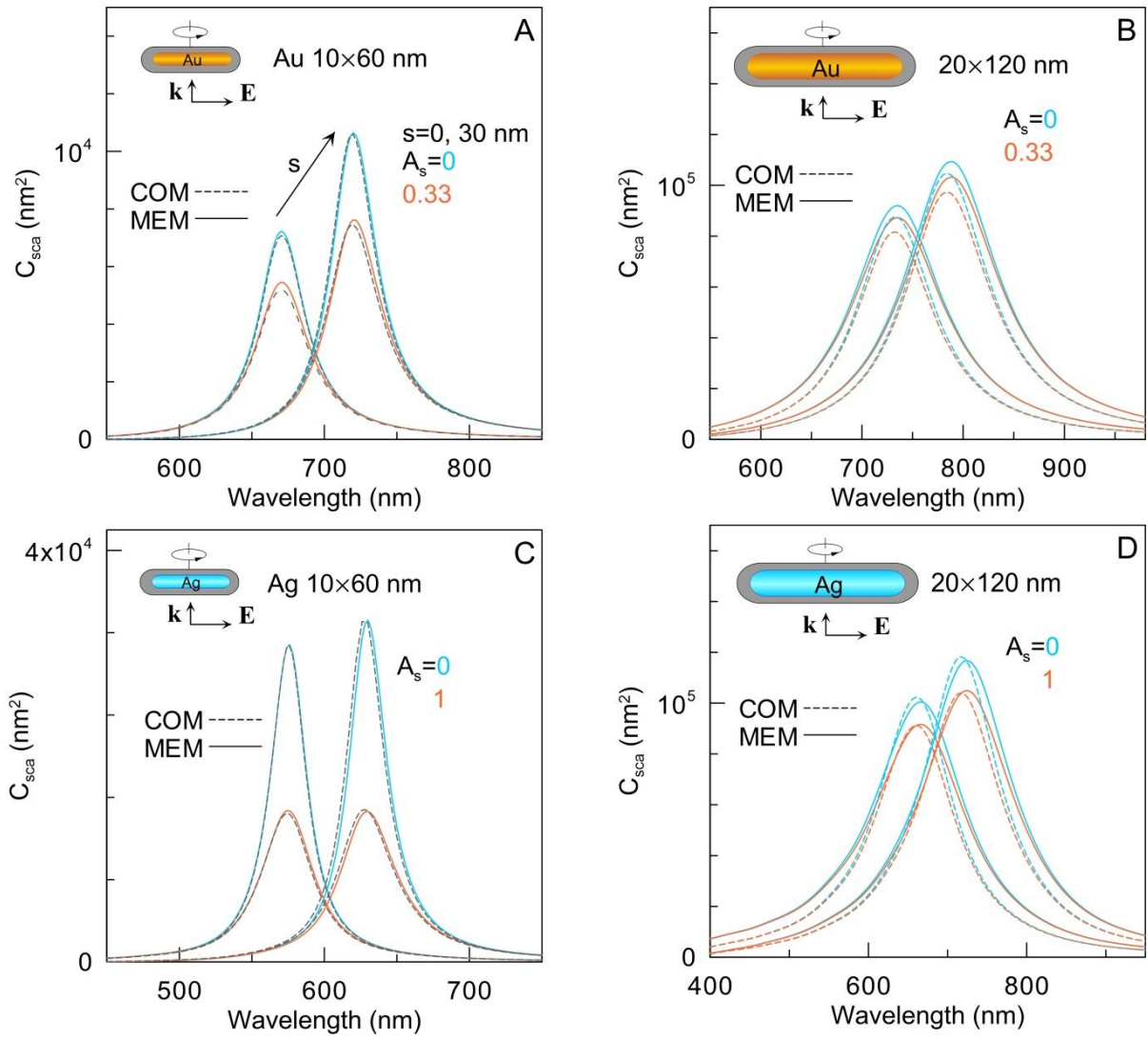

**Figure S4**. Scattering spectra of gold (A, B) and silver (C, D) bare and coated ($s = 30$ nm) nanodisks with dimensions of $10 \times 60$ nm (A, C) and $20 \times 120$ nm (B, D). Calculations by COMSOL (dashed lines) and MEM (solid lines) at in-plane excitation for bulk ($A_s = 0$, blue) and size-corrected optical constants ($A_s = 0.33$ for gold and $A_s = 1$ for silver, orange).

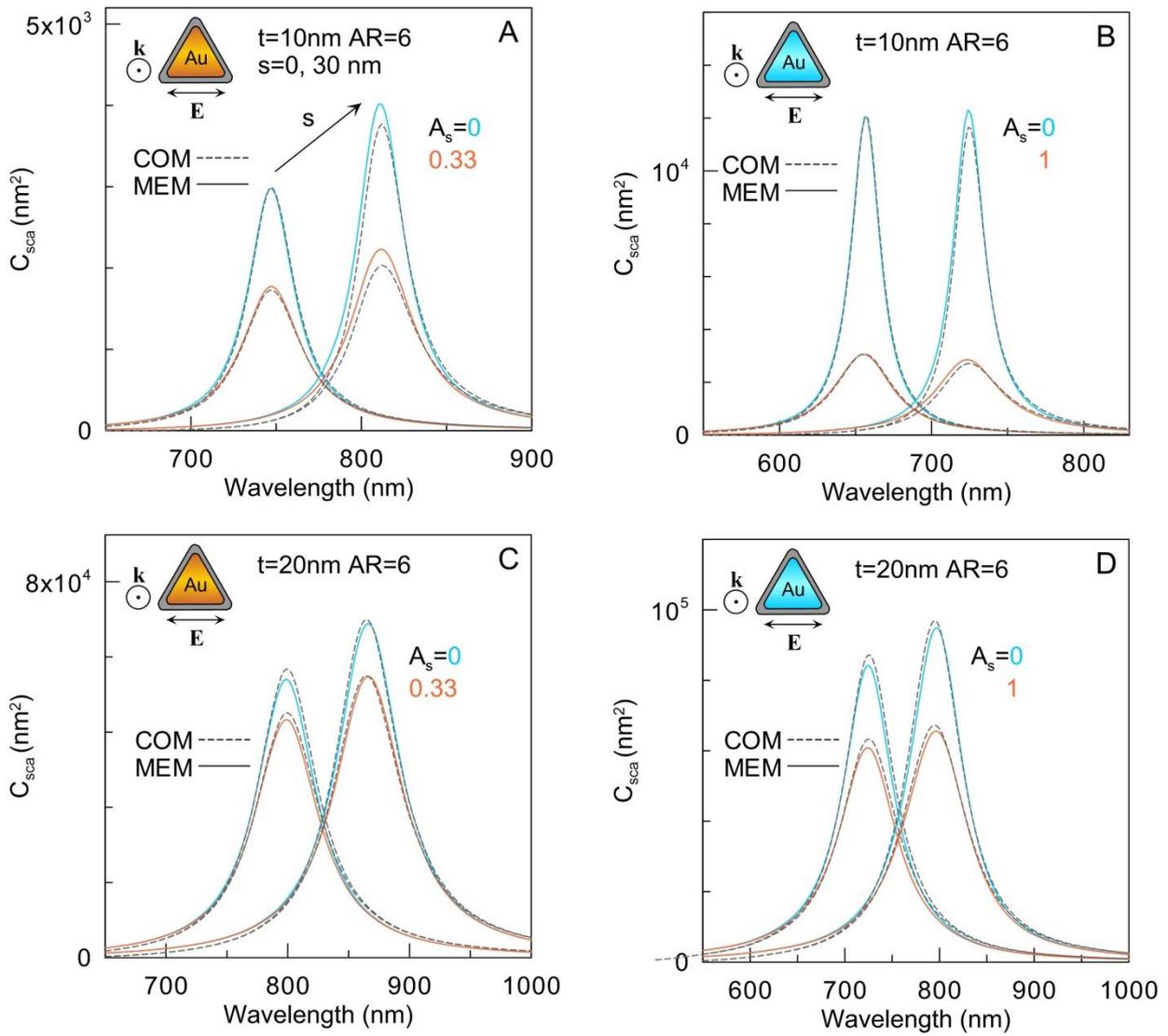

**Figure S5**. Scattering spectra of gold (A, C) and silver (B, D) bare and coated ($s = 30$ nm) nanoprisms with dimensions of $10 \times 60$ nm (A, B) and $20 \times 120$ nm (C, D). Calculations by COMSOL (dashed lines) and MEM (solid lines) under in-plane excitation for bulk ($A_s = 0$, blue) and size-corrected optical constants ($A_s = 0.33$ for gold and $A_s = 1$ for silver, orange).

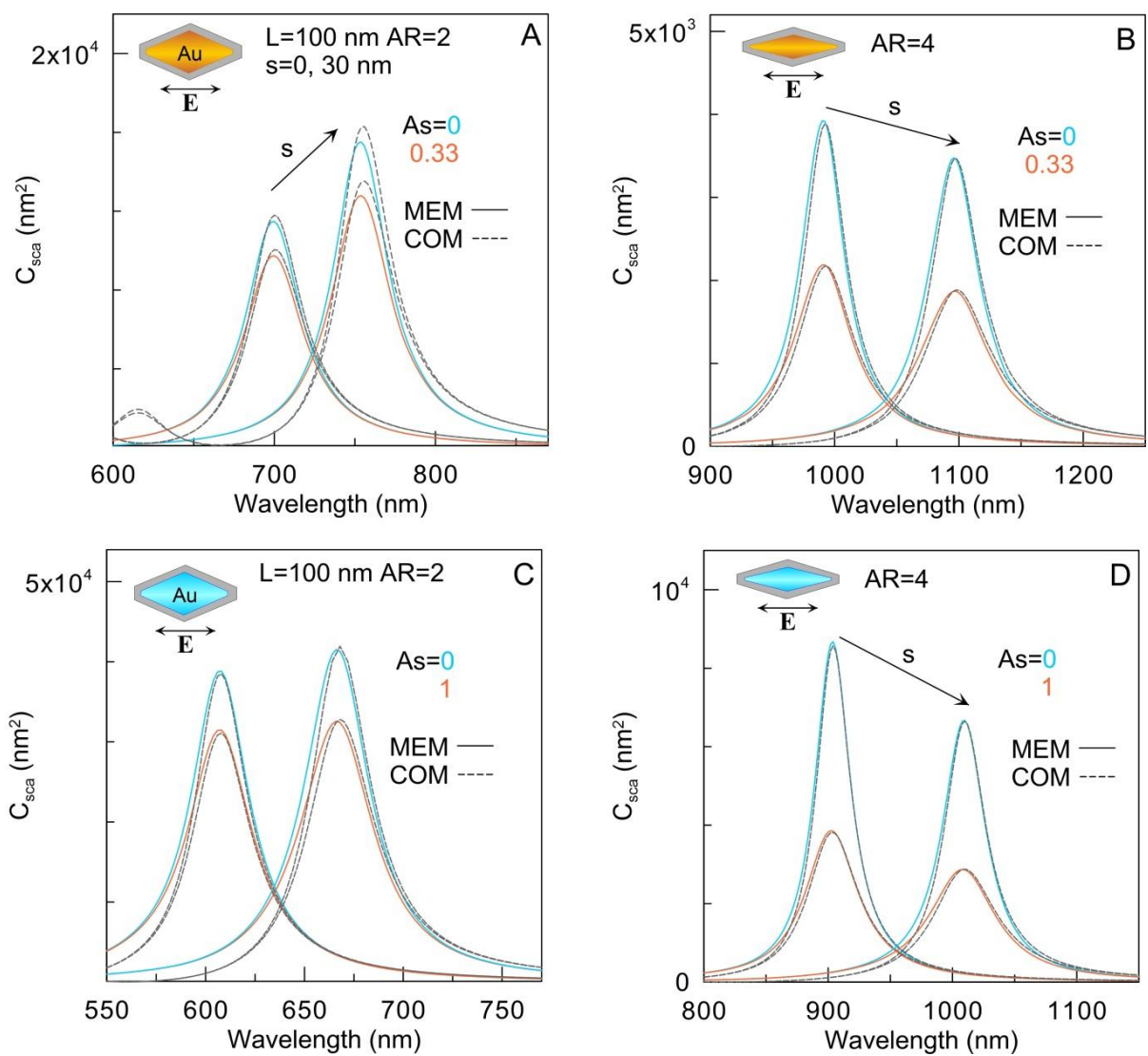

**Figure S6**. Scattering spectra of gold (A, B) and silver (C, D) bare and coated ($s = 30$ nm) bicones with a fixed length of 100 nm and aspect ratios of 2 (A, C) and 4 (B, D). Calculations by COMSOL (dashed lines) and MEM (solid lines) under longitudinal excitation for bulk ($A_s = 0$, blue) and size-corrected optical constants ($A_s = 0.33$ for gold and $A_s = 1$ for silver, orange).

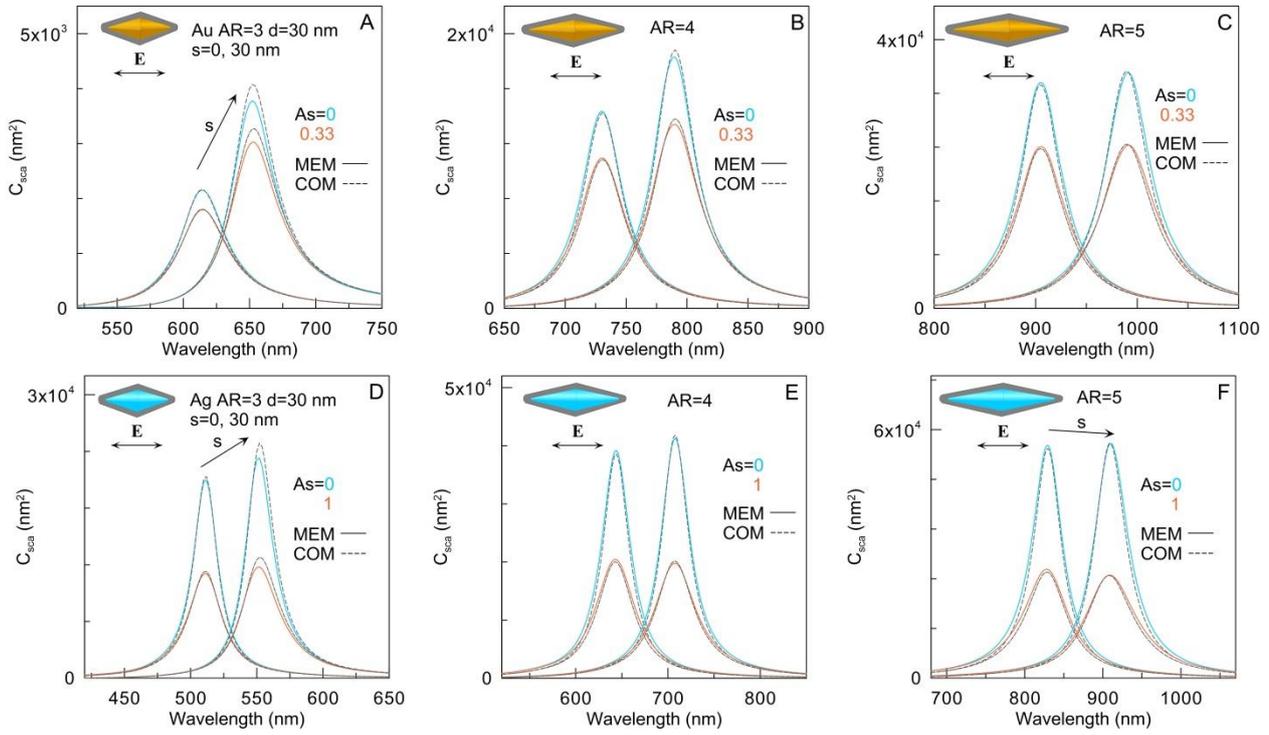

**Figure S7**. Scattering spectra of gold (A–C) and silver (C–D) bare and coated ($s = 30$ nm) bipyramids with a fixed diameter of 30 nm, aspect ratos of 3 (A, D), 4 (B, E) and 5 (C, F), and rounding radii of tips and bases are determined by Eqs. (S23-S24). Calculations by COMSOL (dashed lines) and MEM (solid lines) under longitudinal excitation for bulk ($A_s = 0$, blue) and size-corrected optical constants ($A_s = 0.33$ for gold and $A_s = 1$ for silver, orange).

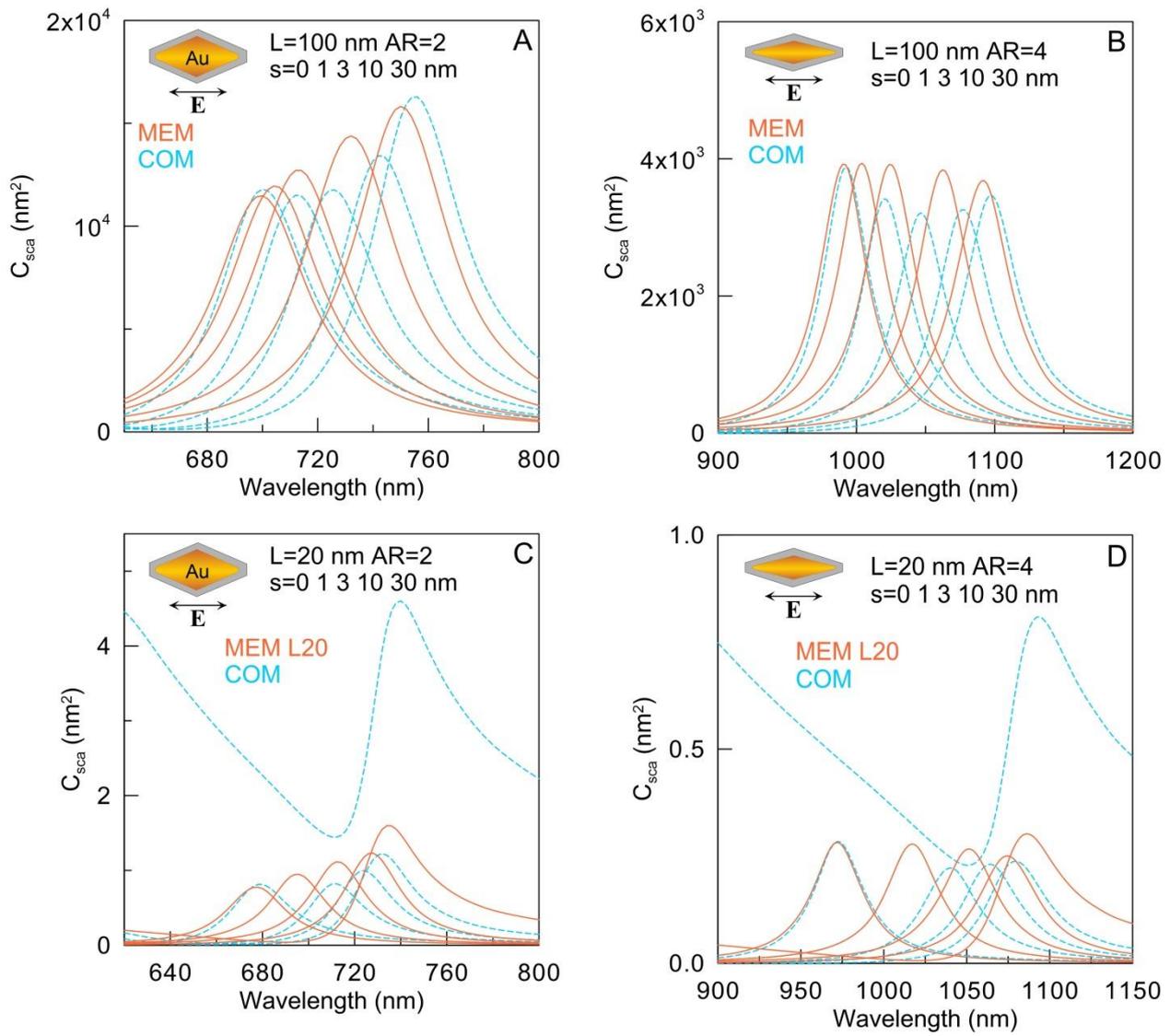

**Figure S8**. Scattering spectra of gold bare ($s=0$) and coated ($s=$ 1, 3, 10, and 30 nm) bicones with a fixed length of 100 nm (A, B) and 20 nm (C, D) and aspect ratios of 2 (A, C) and 4 (B, D). Tip rounding radius is $d/40$ nm. Calculations by COMSOL (blue dashed) and MEM (orange solid lines) under longitudinal excitation for bulk optical constants.

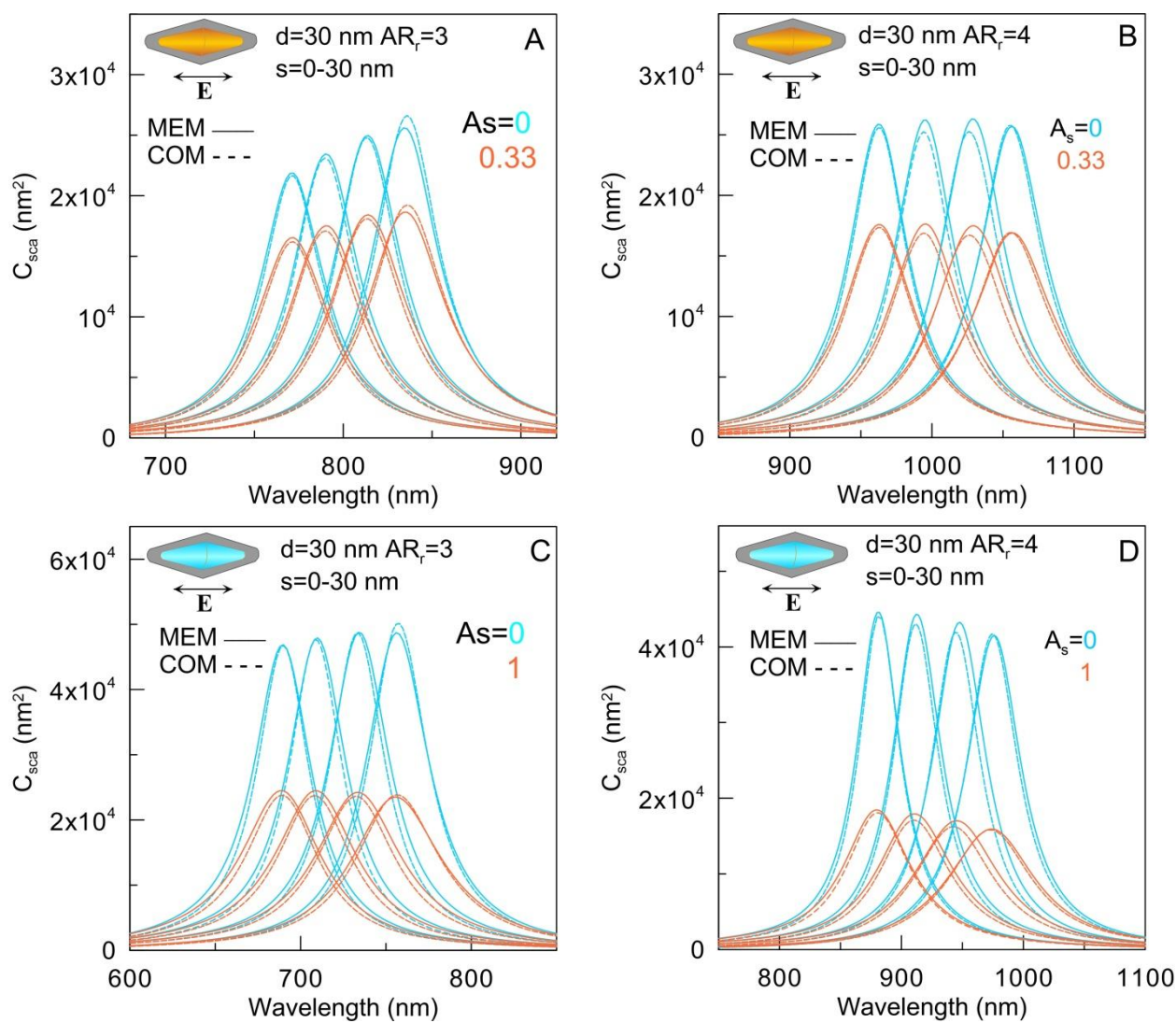

**Figure S9**. Scattering spectra of gold (A, B) and silver (C, D) bare ($s=0$) and coated ($s=3$, 10, and 30 nm) bicones with a fixed diameter of 30 nm, aspect ratios of 3 (A, C) and 4 (B, D), and the tip rounding radius as given by Eq. (S27). Calculations by COMSOL (dashed lines) and MEM (solid lines) under longitudinal excitation for bulk ($A_s = 0$, blue) and size-corrected optical constants ($A_s = 0.33$ for gold and $A_s = 1$ for silver, orange).

# References

[1] R. Yu, L. M. Liz-Marzán, and F. J. García de Abajo, "Universal analytical modeling of plasmonic nanoparticles," Chem. Soc. Rev. 46, 6710–6724 (2017).

[2] N. G. Khlebtsov and S. V. Zarkov, "Combining the modal expansion and dipole equivalence methods for coated plasmonic particles of various shapes," J. Phys. Chem. C 129, 10958-10974 (2025).

[3] J. L. Montaño-Priede, A. Sánchez-Iglesias, S. A. Mezzasalma, J. Sancho-Parramon, M. Grzelczak, "Quantifying shape transition in anisotropic plasmonic nanoparticles through geometric inversion. application to gold bipyramids," J. Phys. Chem. Lett. 15, 3914–3922 (2024).

[4] E. A. Coronado and G. C. Schatz, "Surface plasmon broadening for arbitrary shape nanoparticles: A geometrical probability approach," J. Chem. Phys. 119, 3926–3934 (2003).